\documentclass[onecolumn,a4paper,12pt]{elsarticle}
\pdfoutput=1
\usepackage{ifpdf}
\usepackage{hyperref}
\usepackage[euler]{textgreek}
\usepackage[draft]{minted}

\usepackage{tabto}
\usepackage{multicol}
\usepackage{graphicx}
\usepackage{float}
\usepackage{graphics}
\usepackage{pdflscape}
\usepackage{multirow}
\usepackage{makecell}
\usepackage{adjustbox}
\usepackage{rotating}
\usepackage{paralist}  
\usepackage{longtable}  
\usepackage{comment}  
\usepackage{colortbl}
\usepackage{enumitem}
\usepackage{tikz}
\usepackage{xparse}
\usepackage[utf8]{inputenc}
\usepackage{listings,xcolor}
\usepackage{multicol}
\usepackage{graphicx}
\usepackage{float}
\usepackage{graphics}
\usepackage{pdflscape}
\usepackage{multirow}
\usepackage{makecell}
\usepackage{adjustbox}
\usepackage{rotating}
\usepackage{paralist} 
\usepackage{longtable} 
\usepackage{comment} 
\usepackage{colortbl}
\usepackage{tikz}

\usepackage{tcolorbox}
\usepackage{xparse}

\definecolor{iscte-iul-palette}{HTML}{14BFB8}
\usepackage{xspace}
\usepackage[linesnumbered,ruled,vlined]{algorithm2e}
\usepackage{amssymb}
\usepackage{enumitem}
\usepackage[percent]{overpic}
\usepackage{tablefootnote}

\usepackage{tcolorbox}
\usepackage[position=above, skip=0pt]{caption}

\journal{Computer Standards and Interfaces}

\usepackage{numcompress}

\newcommand{\chapterquotes}[4]{ 
\begin{flushright}
\rightskip=0.8cm\textit{``#1''\footnote{#2}}\\
\vspace{.2em}
\rightskip=.8cm---#3\footnote{#4}
\end{flushright}
\vspace{1em}
}

\NewDocumentCommand{\plot}{m O{} O{12} O{H} O{trim=0cm 0cm 0cm 0cm} }{
\begin{figure}[#4]
\centering
\includegraphics[width=#3cm,clip,#5]{#1}
\caption{#2}
\label{#1}
\end{figure}
}

\newcommand{\noteboard}[2]{
\vspace{0.3cm}
\begin{tcolorbox}[width=\textwidth,colback={white},title={\textbf{#1}},colbacktitle=gray!10,coltitle=black, boxrule=0.5pt, leftrule=3pt, sharp corners=all]  
    #2
\end{tcolorbox}
\vspace{0.3cm}
}

\newcommand{\hypnull}{
\textbf{H\textsubscript{0}}\xspace
}

\newcommand{\hypalt}{
\textbf{H\textsubscript{1}}\xspace
}

\usepackage{tikz}
\usetikzlibrary{svg.path}
\definecolor{orcidlogocol}{HTML}{A6CE39}
\tikzset{
	orcidlogo/.pic={
		\fill[orcidlogocol] svg{M 100 100 C 100 72.4 77.6 50 50 50 C 22.3 50 0 72.4 0 100 C 0 127.6 22.3 150 50 150 C 77.6 150 100 127.6 100 100 Z};
		\fill[white] svg{M 33.666 77.27 L 27.651 77.27 L 27.651 119.103 L 33.666 119.103 L 33.666 100.198 L 33.666 77.27 Z};
		\fill[white] svg{M 42.493 119.103 L 58.742 119.103 C 74.21 119.103 81.007 108.049 81.007 98.167 C 81.007 87.425 72.609 77.23 58.821 77.23 L 42.493 77.23 L 42.493 119.103 Z M 48.509 82.66 L 58.078 82.66 C 71.71 82.66 74.835 93.011 74.835 98.167 C 74.835 106.565 69.484 113.673 57.766 113.673 L 48.509 113.673 L 48.509 82.66 Z};
		\fill[white] svg{M 34.603 127.813 C 34.603 125.665 32.846 123.868 30.658 123.868 C 28.471 123.868 26.713 125.665 26.713 127.813 C 26.713 130.001 28.471 131.758 30.658 131.758 C 32.846 131.758 34.603 129.961 34.603 127.813 Z};
	}
}
\def\orcid#1{%
	\href{https://orcid.org/#1}{%
		\begin{tikzpicture}[baseline]
			\pic[scale=0.1,yshift=-7em] {orcidlogo};
		\end{tikzpicture}
	}%
}

\begin{document}

\begin{frontmatter}

\title{Unveiling process insights from refactoring practices}


\author[ISCTE-IUL]{João Caldeira
\textsuperscript{\orcid{0000-0003-0960-0179}}}
\ead{jcppc@iscte-iul.pt}


\author[ISCTE-IUL]{Fernando Brito e Abreu
\textsuperscript{\orcid{0000-0002-9086-4122}}}
\ead{fba@iscte-iul.pt}

\author[CISUC,Huawei]{Jorge Cardoso \textsuperscript{\orcid{0000-0001-8992-3466}}}
\ead{jcardoso@dei.uc.pt}

\author[ISCTE-IUL]{José Pereira dos Reis
\textsuperscript{\orcid{0000-0002-2505-9565}}}
\ead{jvprs@iscte-iul.pt}

\address[ISCTE-IUL]{Iscte - Instituto Universitário de Lisboa, ISTAR-Iscte, Lisboa, Portugal}
\address[CISUC]{CISUC, Dept. of Informatics Engineering\\ University of Coimbra, Portugal}
\address[Huawei]{Huawei Munich Research Center, Germany}

            
\begin{abstract}

\noindent\textit{\textbf{Context}}:
Software comprehension and maintenance activities, such as refactoring, are said to be negatively impacted by software complexity. The methods used to measure software product and processes complexity have been thoroughly debated in the literature. However, the discernment about the possible links between these two dimensions, particularly on the benefits of using the process perspective, has a long journey ahead. 

\noindent\textit{\textbf{Objective}}:  To improve the understanding of the liaison of developers' activities and software complexity within a refactoring task, namely by evaluating if process metrics gathered from the IDE, using process mining methods and tools, are suitable to accurately classify different refactoring practices and the resulting software complexity.

\noindent\textit{\textbf{Method}}: We mined source code metrics from a software product after a quality improvement task was given in parallel to \textbf{(117)} software developers, organized in \textbf{(71)} teams. Simultaneously, we collected events from their IDE work sessions \textbf{(320)} and used process mining to model their processes and extract the correspondent metrics.

\noindent\textit{\textbf{Results}}: 
Most teams using a plugin for refactoring \textit{\textbf{(JDeodorant)}} reduced software complexity more effectively and with simpler processes than the ones that performed refactoring using only \textbf{\textit{Eclipse}} native features. We were able to find moderate correlations \textbf{($\approx$43\%)} between software cyclomatic complexity and process cyclomatic complexity. Using only process driven metrics, we computed $\approx$30,000 models aiming to predict the type of refactoring method (automatic or manual) teams had used and the expected level of software cyclomatic complexity reduction after their work sessions. The best models found for the refactoring method and cyclomatic complexity level predictions, had an accuracy of \textbf{92.95\%} and \textbf{94.36\%}, respectively.

\noindent\textit{\textbf{Conclusions}}: 
We have demonstrated the feasibility of an approach that allows building cross-cutting analytical models in software projects, such as the one we used for detecting manual or automatic refactoring practices. Events from the development tools and support activities can be collected, transformed, aggregated, and analyzed with fewer privacy concerns or technical constraints than source code-driven metrics. This makes our approach agnostic to programming languages, geographic location, or development practices, making it suitable for challenging contexts such as in modern global software development projects. Initial findings are encouraging, and lead us to suggest practitioners may use our method in other development tasks, such as, defect analysis and unit or integration tests.

\end{abstract}



\end{frontmatter}



\section{Introduction}
\label{sec:introduction}

\chapterquotes{...All things - from the tiniest virus to the greatest galaxy - are, in reality, not things at all, but processes...}{In \textit{"Future Shock"}, Penguin Random House, New York, 1970.}{Alvin Toffler(1928-2016)}{American writer, futurist, and businessman known for his works discussing modern technologies, including the digital and the communication revolutions, with emphasis on their effects on cultures worldwide.}

A process\footnote{Adapted from \href[pdfnewwindow=true]{https://dictionary.cambridge.org/dictionary/english/process}{https://dictionary.cambridge.org/dictionary/english/process}} is \textit{"a series of actions taken in order to achieve a result"}. In many business areas, either on delivering products and/or services, the quality of the outcome is very often related with the process followed to build it \cite{Deming1986OutPosition,Ishikawa1985WhatWay,Taguchi1986IntroductionProcesses}. This is expected to be no different in the software development domain. Therefore, to fully comprehend how software quality and improved maintainability are achieved, one should look carefully to the process perspective to complement any code related analysis \cite{Fuggetta2014SoftwareProcess}.

Software development is intrinsically a process and, accordingly, it is a blend of activities performed by developers, often working from different locations and using a multitude of languages, tools and methodologies in order to create a new product or maintain an existing one \cite{Fuggetta2014SoftwareProcess}. Since the early days of software development, it was understood that programming is an inherently complex and error-prone process, and to fully understand it, we should mine, in a timely and proper manner, all facets of that process \cite{VanderAalst2016}. Any relevant insights one may obtain should therefore originate from the activities and/or artifacts recorded in software repositories during the development life cycle. 

Studies on estimating the effort to develop a certain artifact, the identification of software defects, the prediction of time to solve bugs or on software comprehension, and the detection of refactoring opportunities, are amongst the most common use cases for those repositories \cite{Moha2006AutomaticDefects,Malhotra2018AnalyzingApplications,Karna2019ApplicationProjects,Batarseh2018PredictingAnalytics,Kersten2006UsingProductivity,Murphy2009UsingLab,Negara2013ARefactorings2}.

Refactoring on its own is still a very challenging activity. The identification of components to refactor and the forecast of which methods to embrace continue to be relevant topics for research \cite{Vakilian2011TheData,Vakilian2012UseRefactorings,Ratzinger2007MiningRefactoring,Aniche2020TheRefactoring}. These challenges emerge partially due to the significant functionality limitations software repositories contain and the type of data they use \cite{Negara2012IsEvolution2}.

Some authors confirm that developers perform refactoring tasks manually more frequently than automatically \cite{Negara2013ARefactorings2}. Furthermore, it has been observed, in a real-life scenario, that refactoring can be harmful when done manually, using only IDE native features or simply driven by developers' skills, as it may introduce non-expected defects in the code \cite{Kim2014AnMicrosoft}. 

On trying to comprehend software development processes, including refactoring practices, many data sources, methods, and tools have been used with validated outcomes, but some others are yet to be fully exploited \cite{Menzies2013SoftwareWhat}. For example, since Version Control Systems (VCS) are widely used by developers, researchers get easy access to historical data of many projects and use file-based VCSs as the primary source of code evolution data \cite{Kim2011AnEvolution}. Although it is often convenient to use such repositories, research-based on VCS data is imprecise and incomplete \cite{Negara2012IsEvolution2}. 

As such, answering questions that correlate code changes with other activities (e.g., test runs, refactoring) is often unfeasible. Several reasons may contribute to it, as for instance:
\begin{itemize}
\item developers may not commit all their tests and/or refactorings;
\item there are many ways to refactor one version of the code, therefore it is important to determine the refactoring activities sequences and frequencies;
\item often we cannot distinguish if a refactoring was done manually or through a tool, just by comparing source code snapshots \cite{Murphy-Hill2009HowIt}.
\end{itemize}


\subsection{Code vs. Process Analysis}
Most published work on software quality-related issues is based on source code metrics, especially on Java systems \cite{Finlay2014,Lerthathairat2011AnTechniques,Chen2016ARepositories}. Tools for collecting those metrics upon other frequently used languages, such as JavaScript or Python, are often not available, which expose well the difficulties to reproduce the same research on projects having diverse languages. In case those metric collection tools exist, they often require to share the source code with third-party organizations \cite{Peters2017OnQuality}, particularly on cloud-based platforms. Such scenarios raise privacy and ownership issues on sensitive data. Source code obfuscation does not mitigate this problem because developers need to keep code semantics for interpreting the metrics in context.

Instead, mining the developers' activities and behaviors, the same is to say, to mine their development process fragments, may be a more feasible approach since it is not specific to any programming language, geographic location or development methodology followed.

Event data can be obfuscated without losing the process structure and coherence, therefore, whoever is responsible to analyze the logs can apply algorithms to discover process models in very similar ways as if the logs were not obfuscated \cite{Fahrenkrog-Petersen2020PRIPEL:Information}.
In other words, events from the development tools and support activities can be collected, transformed and aggregated with fewer privacy concerns and technical hurdles. As such, it has been pointed out that software development event logs can be used to complement, or even replace, source code data in software development analytics-related tasks \cite{Casale2016CurrentApplications}.

\subsection{Contributions}

It is frequent to find software prediction models using source code and ownership metrics \cite{Aniche2020TheRefactoring}. However, periodically this data is not easily accessible or has imprecisions. Nowadays, development teams use a diversity of languages, methodologies and tools, therefore, the collection and aggregation of data from software projects remains a challenge. Additionally, process metrics have been found to be good predictors for modeling software development tasks \cite{Rahman2013HowBetter}. 

Thus, we proposed earlier \cite{Caldeira2017} and are now evaluating deeper the use of process metrics gathered from the IDE (Integrated Development Environment), as a way to enhance existing models or eventually, build new ones.

Software product and process metrics have long been proposed, as well as techniques for their collection \cite{Abreu1996MOODMetrics,Watson1996StructuredMetric,Abreu2001MetricsDefinitions,Menzies2013SoftwareWhat,2017MasteringReuse,Cardoso2006,Vanderfeesten2007QualityModels}. However, the association between product and process dimensions is only marginally discussed in the literature \cite{Menzies2018SoftwareNext}. In order to improve our understanding on the liaison between the type of development activities executed and the resulting software product characteristics, namely to ascertain if developers' behavior has an impact on software product quality, we collected data during a software quality improvement task (application of refactoring operations) given to 71 development teams. Regarding developers' behavior, we recorded all events corresponding to the activities/tasks/operations team members performed within their IDE and used those events to mine the underlying process and extract their metrics. Regarding software quality, we collected complexity metrics before and after the refactoring actions took place. The main objectives for this work are, therefore:
\begin{itemize}

    \item[$\bullet$] to assess the use of software process metrics to facilitate and improve the analysis and predictions on refactoring tasks and/or other generic software activities;

    \item[$\bullet$] to evaluate a possible association between the complexity of the produced code and developers' practices in different refactoring tasks;

    \item[$\bullet$] to build classification models for refactoring practices using only process metrics and assess the prediction accuracy of such approach.

\end{itemize}

The rest of this paper is organized as follows: section \ref{sec:Background} provides background related to the research area and emphasizes the need for the followed approach; subsequent section \ref{sec:RelatedWork} outlines the related work; the research methodology and the study setup are presented in section \ref{sec:study-setup}; the results, the corresponding analysis and implications can be found in section \ref{sec:study-results} and threats to validity are discussed in section \ref{sec:Threatsvalidity}; the concluding comments and the outline for future work are produced in section \ref{sec:Conclusions}.

\section{Background}
\label{sec:Background}

Empirical software engineering and software analytics are now mature research areas with substantial contributions to the software development best practices \cite{Zhang2013SoftwarePractice}. The knowledge base created to support those achievements took a great advantage from the experience gathered on analyzing past software projects. Based on the maturity obtained, it was possible to derive several models to measure software complexity, effort and relationships.

\subsection{Early models}

\textbf{Lines of Code(\textit{LOC}).}
The identification and quantification of software \\size/defect relationship did not happen overnight. The first known “size” law, saying the number of defects D was a function of the number of \textit{LOC}; specifically, D = 4.86 + 0.018 * i, was the result of decades of experience and was presented by Fumio Akiyama \cite{Akiyama1971AnDebugging}.\\

\noindent\textbf{Cyclomatic Complexity.}
One of the most relevant propositions to assess the difficulty to maintain software was introduced by Thomas McCabe when he stated that the complexity of the code was more important than the number of \textit{LOC}.
He argued that when his “cyclomatic complexity” metric was over 10, the code is more likely to be defective \cite{Mccabe1976AMeasure}. This metric, underpinned by graph theory, went through thorough validation scrutiny and then became the first software metric recognized by a standardization body, the NIST \cite{Watson1996StructuredMetric}, what makes it even more relevant in the context of this journal.\\

\noindent\textbf{Halstead Complexity.} On trying to establish an empirical science of software development, Maurice Howard Halstead, introduced the Halstead complexity measures \cite{Hariprasad2018SoftwareMetrics}. These metrics, which are computed statically from the code, assume that software measurement should reflect the implementation or expression of algorithms in different languages, but be independent of their execution on a specific platform. Halstead's metrics were used, among other things, to assess programmers' performance in software maintenance activities (measured by the time to locate and successfully correct the bug) \cite{Curtis1979ComplexityMetrics}.\\

\noindent\textbf{Effort Estimators.}
Later, Barry Boehm proposed an estimator for development effort that was exponential on program size: effort = $a * KLOC^{b} * EffortMultipliers$, where 2.4 $\leq$ a $\leq$ 3 and 1.05 $\leq$ b $\leq$ 1.2 \cite{Boehm1981SoftwareEconomics}.\\

\noindent\textbf{Henry and Kafura Metrics.}
These two authors defined and validated a set of software metrics based on the measurement of information flow between system components. Specific metrics are defined for procedure complexity, software modules complexity, and module coupling \cite{Henry1981SoftwareFlow}.\\

The above models were the foundation knowledge for what is nowadays often categorized as Software Development Analytics \cite{Buse2010AnalyticsDevelopment}. However, current development methods, tools and data repositories are very different from the past. Back in those years, software developers were mainly using a text editor and a compiler. Software projects were essentially built employing a single programming language, following a fairly simple development methodology and the developers were rarely located in different geographies or across multiple time zones. These workspace conditions have changed.

\subsection{Modern Days}

In 2019, JetBrains\footnote{\href[pdfnewwindow=true]{https://www.jetbrains.com/lp/devecosystem-2019/}{https://www.jetbrains.com/lp/devecosystem-2019/}} polled almost 7000 developers about their development ecosystem. Results show that more than 30 different programming languages are being used and confirmed that web back-end, web front-end and mobile applications are the type of applications mostly developed, with figures of 60\%, 46\% and 23\%, respectively. It was unanimous the adherence of cross-platform development frameworks and 80\% said they use any type of source code collaboration tool, 75\% use a standalone IDE and 71\% use a lightweight desktop editor. Almost 50\% said they use continuous integration/delivery (CI/CD) and issue tracking tools. Less than 15\% responded that they use any sort of static analysis, code review and in-cloud IDE tools. Table \ref{table:survey-key-takeaways} presents the key takeaways from the mentioned survey.

\begin{table}[t]
\footnotesize
\caption{Survey Key Takeaways*}
\label{table:survey-key-takeaways}
\begin{tabular}{lll}
	\hline\noalign{\smallskip}
	\textbf{Findings} & \textbf{} & \textbf{}\\
	\noalign{\smallskip}\hline\noalign{\smallskip}

 \\
\multicolumn{3}{l}{\cellcolor{gray!10}\textbf{Programming Languages Overall Results}}\\[0.1cm]
Java & The most popular primary programming language &\\[0.1cm]
JavaScript & The most used overall programming language &\\[0.1cm]
Go & The most promising language as 13\% said they will adopt it &\\[0.1cm]
Python & Most studied language as 27\% said they used it in the last 12 months &\\[0.4cm]

\multicolumn{3}{l}{\cellcolor{gray!10}\textbf{Languages used in last 12 months}}\\[0.1cm]
\multicolumn{3}{l}{JavaScript(69\%), HTML/CSS(61\%), SQL(56\%), Java(50\%), Python(49\%)}\\[0.1cm]
\multicolumn{3}{l}{Shell Scripting(40\%), PHP(29\%), TypeScript(25\%), C\#(24\%), C++(20\%)}\\[0.3cm]

\multicolumn{3}{l}{\cellcolor{gray!10}\textbf{Development Environments(Operating Systems)}}\\[0.1cm]
\multicolumn{3}{l}{Windows(57\%), macOS(49\%), Unix/Linux(48\%), Other(1\%)}\\[0.3cm]

\multicolumn{3}{l}{\cellcolor{gray!10}\textbf{Type of Application Development}}\\[0.1cm]
\multicolumn{3}{l}{Web Back-End(60\%), Web Front-End(46\%), Mobile(23\%), Libraries(14\%)}\\[0.1cm]
\multicolumn{3}{l}{Desktop(12\%), Other Back-End(16\%), Data Analysis(13\%), Machine Learning(7\%)}\\[0.3cm]

\multicolumn{3}{l}{\cellcolor{gray!10}\textbf{Type of Tests Used}}\\[0.1cm]
\multicolumn{3}{l}{Unitary(71\%), Integration(47\%), End-to-End(32\%), Other(2\%), Don't Test(16\%)}\\[0.3cm]

\multicolumn{3}{l}{\cellcolor{gray!10}\textbf{Targeted Mobile Operating Systems \& Frameworks Used}}\\[0.1cm]
\multicolumn{3}{l}{Android(83\%), iOS(59\%), Other(3\%)}\\[0.2cm]

\multicolumn{3}{l}{React Native(42\%), Flutter(30\%), Cordova(29\%), Ionic(28\%), Xamarin(26\%)}\\[0.3cm]

\multicolumn{3}{l}{\cellcolor{gray!10}\textbf{Regularly Used Tools}}\\[0.1cm]
\multicolumn{3}{l}{Source Code Collaboration Tool(80\%), Standalone IDE(75\%)}\\[0.1cm]
\multicolumn{3}{l}{Lightweight Desktop Editor(71\%), CI/CD Tool(45\%), Issue Tracker(44\%)}\\[0.1cm]
\multicolumn{3}{l}{Static Analysis Tool(13\%), Code Review Tool(10\%)}\\[0.1cm]

   \noalign{\smallskip}\hline
   \multicolumn{3}{l}{\shortstack[l]{*All values(\%) represent the percentage of affirmative respondents}}
\end{tabular}
\end{table}

In summary, currently, a software development ecosystem has to deal with at least the following facets: 
\begin{itemize}
    \item[$\bullet$] \noindent\textbf{Multi-Language Ecosystem.} According to a recent work about multi-language software development \cite{Mayer2017OnDevelopers}, the authors present evidences that non-trivial enterprise software systems are written in at least 7 programming languages and, a previous work showed that in the open source world alone, the average is 5 languages per project. Among these, one may find general-purpose languages(GPL) such as Java or C\# and also domain-specific languages(DSL) like SQL and HTML, and cross-language links are also quite common, meaning some code artifacts are shared between languages. As a result, developers confirm they find more problems in activities such as implementing new requirements (78\%) and in refactoring (71\%).
    
    \item[$\bullet$] \noindent\textbf{IDE Evolution.} A substantial change was carried in the Integrated Development Environments (IDEs). Software development moved away from the early days of the code editor. As confirmed by the Jetbrains poll, developers now use powerful platforms and frameworks which allow them to be more productive on their jobs. This results from the combination of different software development life cycle activities, such as: requirements elicitation, producing analysis and design models, programming, testing, configuration management, dependencies management or continuous integration into one single tool such as \texttt{Eclipse}, \texttt{IntelliJ IDEA}, \texttt{Netbeans} or \texttt{Visual Studio Code}. These tools support the needs of different stakeholders, as they embed a myriad of plugins available in their marketplaces. These plugins are not just available, they are properly customized for specific users/purposes, such as for modellers, programmers, testers, integrators or language engineers.

    \item[$\bullet$] \noindent\textbf{Low Code and No Code Paradigms.} Modern software development practices make consistent use of both approaches. They enable faster development cycles requiring little to no coding in order to build and deliver applications and processes. Low-code development platforms are seen as advanced IDEs which employ drag-and-drop software components and visual interfaces to replace extensive coding. With high-level visual modeling languages, they provide higher levels of abstraction that allow a major reduction in hand-coding to develop an application \cite{Henriques2018ImprovingLanguage}. In the extreme case we have no-code development where, by definition, textual programming is banned, giving rise to the so-called \textit{citizen developers}. The most notable examples are online application generators (OAGs) that automate mobile and web app development, distribution, and maintenance, but this approach is claimed to be pledged with security vulnerabilities \cite{Oltrogge2018TheGenerators}. This paradigm shift in software development may also require a change in the way we assess critical properties of a software project, such as, quality, maintainability, and evolvability.
    
    \item[$\bullet$] \noindent\textbf{Global Software Development.} The aforementioned IDE platforms facilitated collaboration and the adoption of Global Software Development (GSD). Nowadays, a single software project often has developers, testers and managers located in different time zones and distinct world regions or countries \cite{Niazi2016}.
    
    \item[$\bullet$] \noindent\textbf{CI/CD and DevOps.}
    Continuous Integration and Continuous Deployment (CI/CD) have seen an incremental usage in the last few years. However, efficient CI/CD pipelines are rare, particularly in the mobile apps world where developers seem to prefer the execution of \textit{ad hoc} tasks  \cite{Cruz2019ToApp}.
    Whilst CI/CD focuses more on the automation of tools along a defined software life cycle, DevOps has major concerns with the responsiveness, responsibilities and processes within the development, the deployment and the operational phases of software projects. Keeping these intertwined processes compliant with organizational rules is therefore a persistent requirement.
    
    \item[$\bullet$] \noindent\textbf{Resource Coordination.} It is still one of the fundamental problems in software engineering \cite{Herbsleb2016BuildingAward} and it can be characterized as a socio-technical phenomenon. Understanding the dependencies between development tasks and discover teams' behaviours continues to be a challenge in resource allocation and coordination of modern software projects.

\end{itemize}

Software product repositories have many limitations in terms of the process data they handle. For example, these repositories usually deal only with source code and do not track the developers' geographic location, their workflows within the IDE nor the developers' environment characteristics. A complete repository of process related data with the communications, activities, decisions and actions taken by developers, testers and project managers, are, most of the time, if not always, neglected when the goal is to study a development process. Usually, even if the authors claim they are studying a process, they are often doing it using only artifact related data \cite{Menzies2018SoftwareNext}.

With the existing diversity of languages, methodologies, tools and the fact that resources are now distributed across the world and originate from multiple cultures with different skills, it is somewhat an anachronism to keep using old methods to assess, for example, complexity or build cross-cutting analytical models in current software projects. New approaches, supporting multi-languages, being multi-process aware, and keeping geography diversity transparent are called for, such as our pioneering approach for mining of software development processes based on the IDE event logs. That approach, dubbed Software Development Process Mining \cite{Caldeira2017}, allows reversing engineer a complete software development process, just a process fragment or simply \textit{ad hoc} activities performed by developers, by mining event logs taken from real software development activities.

\section{Related Work}
\label{sec:RelatedWork}

To address the incompleteness of data sources related with software repositories, we strongly believe that Software Development Process Mining based at least on the IDE(but not limited to) can play that role and Process Mining tools and methods can be the vehicles to achieve that goal. Many authors have followed similar paths, bringing not only evidences for its usefulness but also valid contributions to improve established methods.

A decade ago, \cite{Poncin2011a} mined software repositories to extract knowledge about the underlying software processes, and \cite{Rubin2014,Rubin2014b} have learned about user behavior from software at operations. \cite{Ioannou20182} was able to extract events from \texttt{Eclipse} and have discovered, using a process mining tool, basic developers' workflows. Some statistics were computed based on the activities executed and artifacts edited. 

\cite{Mittal2014} presented an application of mining three software repositories: team wiki (used during requirement engineering), version control system (development and maintenance) and issue tracking system (corrective and adaptive maintenance) in the context of an undergraduate Software Engineering course. Experimentation revealed that not only product but process quality varies significantly between student teams and mining process aspects can help the instructor in giving directed and specific feedback. However, in this case, IDE usage mining was not contemplated.

The working habits and challenges of mobile software developers with respect to testing were investigated by \cite{Cruz2019ToApp}. A key finding of this exhaustive study, using 1000 Android apps, demonstrates that mobile apps are still tested in a very ad hoc way, if tested at all. A another relevant finding of this study is that Continuous Integration and Continuous Deployment (CI/CD) pipelines are rare in the mobile apps world (only 26\% of the apps are developed in projects employing CI/CD) - authors argue that one of the main reasons is due to the lack of exhaustive and automatic testing. Therefore, distinguishing during development sessions the type of tests being done can contribute to the overall software quality.

\cite{Yan2019CharacterizingCommits} explored if one can characterize and identify which commits will be reverted.
An identification model (e.g., random forest) was built and evaluated on an empirical study on ten open source projects including a total of 125,241 commits. The findings show that the 'developer' is the most determinant dimension of features for the identification of reverted commits. This suggests that assessing developers behaviors can lead to better understand software products quality.

\cite{Hassan2018StudyingStore} studied the dialogue between users and developers of free apps in the Google Play Store. Evidences found, showed that it can be worthwhile for app owners to respond to reviews, as responding may lead to an increase in the given rating and that studying the dialogue between user and developer can provide valuable insights which may lead to improvements in the app store and the user support process. We believe the same rationale may be applied to comprehend the workflows and dialogues between developers and project owners, and how that may impact software products.

Development activities were extracted by \cite{Bao2018} from non-instrumented applications and used machine learning algorithms to infer a set of basic development tasks. However, in this case, no process mining techniques were used to discover any pattern of application usage. The extraction of usage smells was the focus of 
\cite{Damevski2017}, where a semi-automatic approach was adopted to analyze a large dataset of IDE interactions using cluster analysis. Again, process mining techniques were not used.
Process mining was indeed used by \cite{Leemans2018} to gain knowledge on software under operation (not under development) by analyzing the hierarchical events produced by application calls(eg: execution of methods within classes) at runtime. 

\cite{Xia2018MeasuringProfessionals} collected events from the IDE to measure program comprehension and evaluated the correlation between developers' activities and the time they spent on them. Despite the fact that a process was being studied, no evidence of using process mining methods was provided.

A few authors have also followed the route we suggested earlier and resumed in \cite{Caldeira2019AssessingMining}. As such, we are witnessing more evidences that it is indeed a valid approach, therefore, \cite{Ardimento2019EvaluatingApproach} used process mining to evaluate developers' coding behavior in software development processes. Process models were discovered and used to classify the developers as low-performing and high-performing profiles. With a similar goal, in \cite{Ardimento2019LearningApproach}, a different miner algorithm was assessed to obtain complementary results and in \cite{Ardimento2019MiningLogs}, developers' profiling was achieved by mining event logs from a web-based cloud IDE.

Finally, \cite{Aniche2020TheRefactoring} highlights the importance of having more fine-grained process metrics in prediction models and evaluated several machine learning algorithms in predicting software refactoring opportunities. This work focuses on deciding when, what and why to refactor, however, it does not address which refactor practice was indeed applied.

The studies mentioned above used a multitude of process mining techniques, statistics and machine learning methods. Different data source types have been used to extract the information needed to support them. However, to the best of our knowledge, none of these works combine process and product metrics with the aim of assessing potential correlations and/or impacts between the process and the product. Moreover, none uses only process metrics to discover work patterns or to predict development behaviors, particularly, refactoring practices.


\section{Study Setup}
\label{sec:study-setup}

We setup an environment where the same well-defined tasks on software quality assurance was performed independently by several teams.

Our research guaranteed that all teams had similar backgrounds and performed the same task upon the same software system. This approach was used to block additional confounding factors in our analysis. The task targeted a complex open-source Java system named \texttt{Jasml (Java Assembling Language)}\footnote{\href[pdfnewwindow=true]{http://jasml.sourceforge.net/}{http://jasml.sourceforge.net/}}.

To understand the work developed by each team in each task, we collected the corresponding IDE events for mining the underlying process. At the end of each task, we also collected the modified \texttt{Jasml} project code for each team and obtained the corresponding product metrics.

\subsection{Subject Selection}

Our subjects were the finalists (3rd year) of a B.Sc. degree on computer science at the ISCTE-IUL university, attending a compulsory software engineering course. They had similar backgrounds as they have been trained across the same set of courses along their academic path. Teams were assembled with up to 4 members each and were requested to complete a code smells detection assignment, aiming at identifying refactoring opportunities and then to apply them.

\subsection{Data Collection}
The participants were requested to perform the refactoring tasks in two different ways: \textbf{Automatically} and \textbf{Manually}. 

The refactoring tasks had the following requirements:
\begin{itemize}

\item \textbf{Automatic Refactoring(AR).} This task was executed from March 1\textsuperscript{st} to March 20\textsuperscript{th}, using \texttt{JDeodorant}\footnote{\href[pdfnewwindow=true]{https://users.encs.concordia.ca/~nikolaos/jdeodorant/}{https://users.encs.concordia.ca/~nikolaos/jdeodorant/}}. This tool suggests refactoring opportunities by detecting, among others, the following four well-known code smells: \texttt{Long Method, God Class, Feature Envy} and \texttt{Type Checking} \cite{Beck1999BadCode}. Once participants have detected the occurrences of those code smells, they were required to apply \texttt{JDeodorant}'s fully automated refactoring features to fix the critical ones. 

\item \textbf{Manual Refactoring(MR).} This task was pursued from March 21\textsuperscript{st} to 28\textsuperscript{th} and differed from the previous one because \texttt{JDeodorant} automatic refactoring capabilities were banned. Instead, subjects could use Eclipse's native interactive refactoring features or perform the refactorings manually.\\

\end{itemize}

The \texttt{Eclipse} IDE has an internal event bus accessed by the interface\\ \texttt{IEventBroker}\footnote{\href[pdfnewwindow=true]{https://wiki.eclipse.org/Eclipse4/RCP/Event\_Model}{https://wiki.eclipse.org/Eclipse4/RCP/Event\_Model}} which is instantiated once the application starts. It contains a publishing service to put data in the bus, whilst the subscriber service reads what's in that bus. 
Using this feature we developed an \texttt{Eclipse} plugin\footnote{\href[pdfnewwindow=true]{https://github.com/jcaldeir/iscte-analytics-plugins-repository}{https://github.com/jcaldeir/iscte-analytics-plugins-repository}} capable of listening to the actions developers were executing. Before the experiment, the plugin was installed on each subject's IDE, and later, all subjects received an unique \textit{username/key} pair as credentials. 

\newpage
A sample event instance collected with our plugin is presented in listing \ref{lst:eclipse_event}. The field tags are self explanatory.


\lstset{
    string=[s]{"}{"}, stringstyle=\color{black},
    comment=[l]{:},  commentstyle=\color{iscte-iul-palette}
}
\begin{lstlisting}[basicstyle=\scriptsize, caption=Sample Eclipse Event Instance in \texttt{JSON} format, label=lst:eclipse_event]
{ 
"team" : "Team-10",
"session"  : "dkoep74-ajodje5-63j3k2",
"timestamp_begin" : "2019-05-03 16:53:52.144",
"timestamp_end" : "2019-05-03 16:54:04.468",
"fullname"  : "John User",
"username"  : "john",
"workspacename"  : "Workspace1",
"projectname"  : "/jasml_0.10",
"filename"  : "/jasml_0.10/src/jasml.java",
"extension"  : "java",
"categoryName": "Eclipse Editor",
"commandName": "File Editing",
"categoryID": "org.eclipse.ui.internal.EditorReference",
"commandID": "iscte.plugin.eclipse.commands.file.edit",
"platform_branch": "Eclipse Oxygen",
"platform_version": "4.7.3.M20180330-0640",
"java": "1.8.0_171-b11",
"continent": "Europe",
"country": "Portugal",
"city": "Lisbon",
....
"hash": "00b7c0ef94e02eb5138d33daf38054e3" //To detect event tampering
}
\end{lstlisting}

\subsubsection{Data Storage}
Collected data was stored locally on each subject's computer in a CSV file. Whenever Internet connection was available, the same data was stored in real-time in the cloud\footnote{\href[pdfnewwindow=true]{https://azure.microsoft.com/en-us/services/event-hubs/}{https://azure.microsoft.com/en-us/services/event-hubs/}}. This storage replication mechanism allowed for offline and online collection\footnote{The plugin currently supports the collection of events locally in CSV and JSON files; stream events to Azure Event Hub and Kafka remotely; and uses an integration with Trello to extract project activities which can be triggered as manual events by the developers. Kafka and Trello integrations were not used in this experiment.}. The final dataset, combining the two different sources, was then loaded into a MySQL database table where the username and event timestamps that formed the table's unique key were used to detect and avoid duplicated data insertions. Figure \ref{chapter3-0-1.pdf} presents the complete schema for the data collection workflow. We use the BPMN standard process definition language for that purpose \cite{chinosi2012bpmn}.

\plot{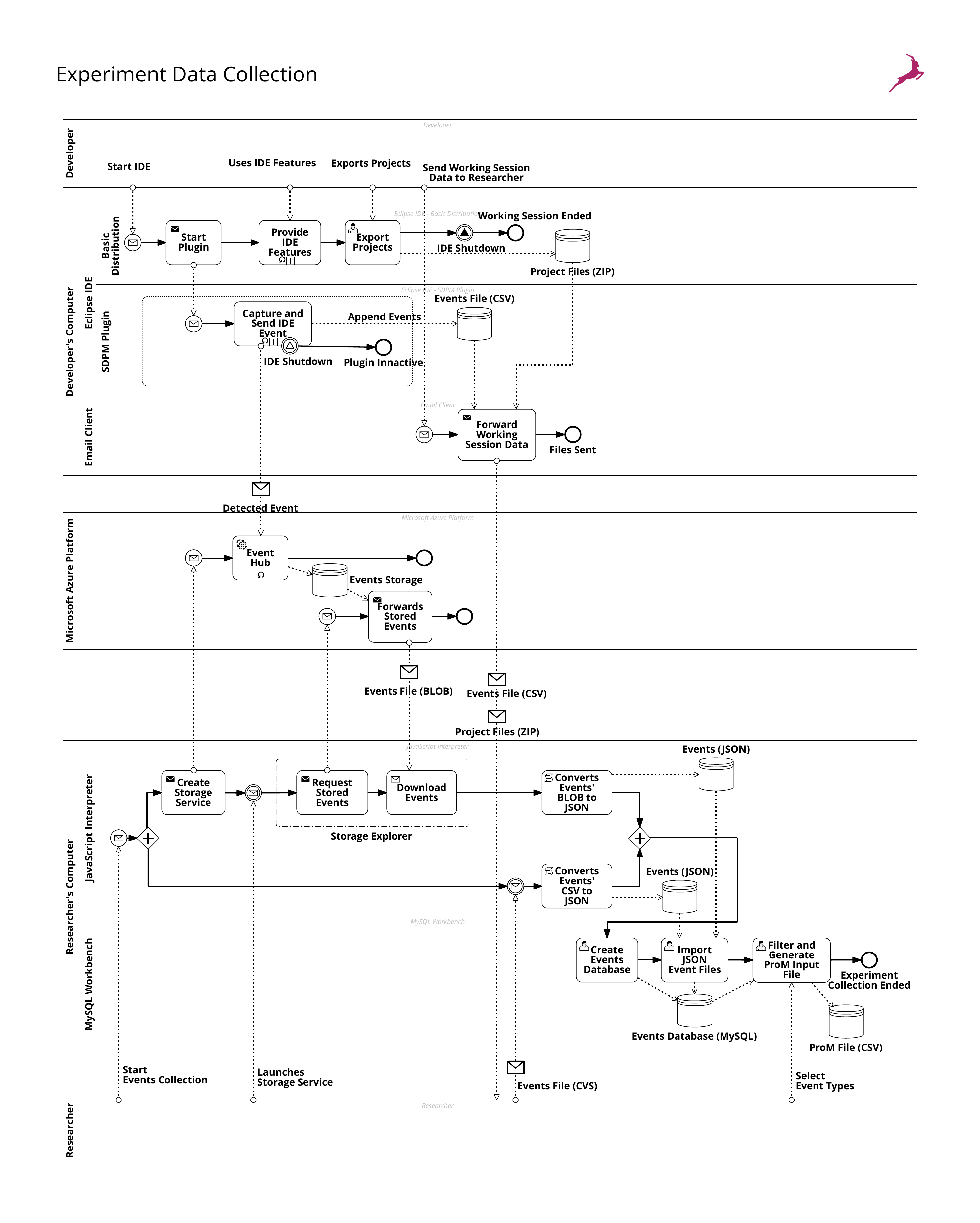}[Experiment Data Collection Workflow][13.5][ht!][trim=1.9cm 1.5cm 1.9cm 3.6cm]

\subsubsection{Data Preparation}

When the software quality task ended, we collected from each team their projects' code together with the events files containing the actions performed during the aforementioned activities. As such, each team produced and delivered two new \texttt{Jasml} projects, one for the automatic and another for the manual refactoring. The events files would map events for the two different tasks, as they were done in different time frames.

All events stored in the database were imported into the ProM process mining tool\footnote{Version 6.8, available at \href[pdfnewwindow=true]{http://www.promtools.org}{http://www.promtools.org}} and converted to the IEEE eXtensible Event Stream (XES) standard format \cite{Gunther2014XesDefinition}. The following event properties were mapped when converting to XES format:
\begin{itemize}
  \item \textit{team} and \textit{session} were used as \textbf{CaseID} since we were interested to look into process instances of teams and their multiple development sessions, not of individual programmers.
  \item Properties \textit{filename}, \textit{categoryName} and \textit{commandName} forming a hierarchical structure were used as the \textbf{Activity} in the process.
  \item The \textit{timestamp\_begin} and \textit{timestamp\_end} were both used as activity \textbf{Timestamps}.
  \item Other properties were not used in the process discovery phase, however, they were later used for metrics aggregation and context analysis.
\end{itemize}

\subsection{Data Analysis}
\label{data_analysis}

\subsubsection{Context}

All teams started with the same version of \textbf{Jasml 0.10}, therefore, we had two relevant moments to get measures from:

\begin{enumerate} 
\item The initial moment \textbf{\textit{(t0)}}, when we extracted the metrics for the initial product version. However, we didn't know how it was built, therefore, we were missing\footnote{In reality we may consider all of them to be zero} the process metrics.\\
\item The end of the task \textbf{\textit{(t1)}}, when we extracted again the product metrics for the changed \textbf{Jasml 0.10} project of each team as they stand after the refactoring sessions. In addition, we had also IDE usage events which provide evidences on how the product was changed.
\end{enumerate}

Following data extraction, we computed, for each product metric defined in Table \ref{table:product-metrics-description}, their relative variance as shown by Equation \ref{equ:product-delta}. The relative variance variables were the ones we used in all \textbf{RQs}.

\footnotesize
\begin{equation}
\label{equ:product-delta}
\Delta product\ metrics\textsubscript{\textbf{(t1-t0)}} =  \frac{product\ metrics\textsubscript{\textbf{(t1)}} - product\ metrics\textsubscript{\textbf{(t0)}}}{product\ metrics\textsubscript{\textbf{(t0)}}} * 100 \\
\end{equation}
\normalsize

The relative variance was used in order to generalize our approach, thus, making it applicable in scenarios where different teams work on distinct software projects.

Process metrics described in Table \ref{table:process-metrics-description} were derived from the events dataset captured between moments \textbf{\textit{(t0)}} and \textbf{\textit{(t1)}}, either by summing the events or using the method described in \ref{sec:process-metrics}. These metrics may be seen as a representation of the effort done by each team during the refactoring practices.


The complete workflow followed in data pre-processing, aggregation and analysis is presented in Figure \ref{chapter3-0-2.pdf}.


\plot{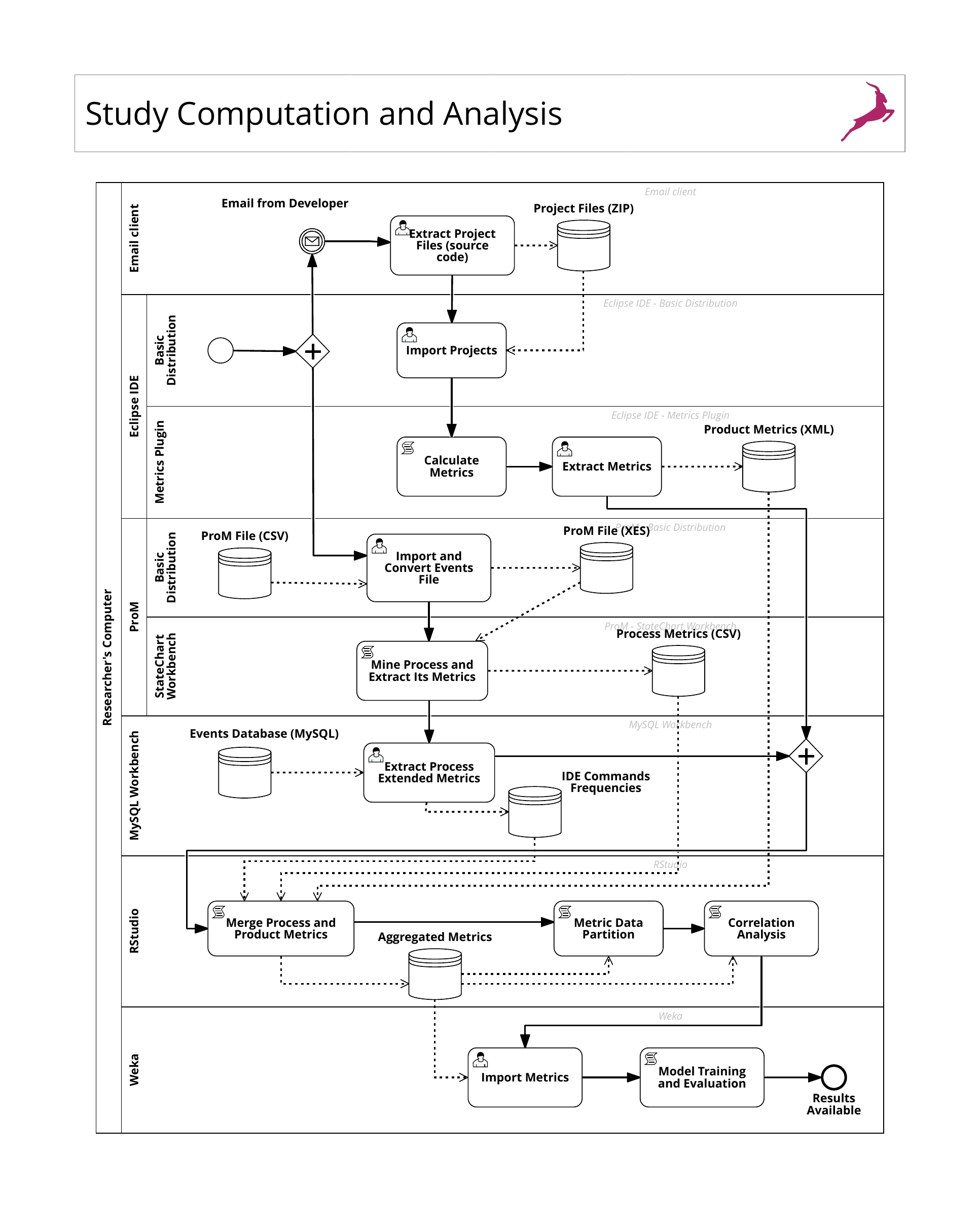}[Study Computation and Analysis Process][13.5][ht!][trim=1.9cm 1.8cm 1.9cm 3.6cm]

\subsubsection{Product and Process Metrics}
\label{sec:product-metrics}
To extract software metrics we used the plugin built by Sauer\footnote{\href[pdfnewwindow=true]{http://metrics.sourceforge.net}{http://metrics.sourceforge.net}}.
Although having more than a decade of age, it is still one of the more proven and popular options regarding open source metrics plugins for Eclipse. 

The plugin itself offers a simple interface and reporting capabilities with which users can define optimal ranges and issue warnings for certain metrics, as well as being able to export calculated metrics to XML files.
The set of metrics obtained by this plugin are presented in Table \ref{table:product-metrics-description} on \ref{sec:AppendixA1}.



\subsubsection{Process Discovery}
\label{sec:process-metrics}
Several well known algorithms exist to discover process models, such as, the $\alpha$-algorithm, the heuristics, genetic and the fuzzy miner amongst others \cite{Maita2018,Garcia2019ProcessStudy}. Our need to discover and visualize the processes in multiple ways lead us to choose the \texttt{ProM's StateChart Workbench} plugin \cite{Leemans2018}. This plugin, besides supporting process model discovery using multiple hierarchies and classifiers, also allows to visualize the model as a Sequence Diagram and use notations such as Petri Nets and Process Trees. This plugin is particularly suitable for mining software logs, where an event structure is supposed to exist, but it also supports the mining of other so-called generic logs.

Events collected from software in operation (e.g. Java programs) reveals the presence of a hierarchical structure, where methods reside within classes, and classes within packages \cite{Leemans2018b}. The same applies to IDE usage actions where identified menu options and executed commands belong to a specific category of command options built-in the \texttt{Eclipse} framework. Supported by this evidence, we used the Software log Hierarchical discovery method with a Structured Names heuristic to discover the processes based on the fact that the events were using a \textit{filename$|$category$|$command} structure (e.g. \texttt{/jasml0.10/src/jasml.java$|$Eclipse Editor$|$File Open}).

Several perspectives can be used to discover and analyze a business process. The commonly used are: \texttt{Control-Flow, Organizational, Social} and \texttt{Performance}. We have focused on the \texttt{Control-Flow} perspective in this paper. It defines an approach that consists in analyzing how each task/activity follows each other in an event log, and infer a possible model for the behavior captured in the observed process.

Process metrics, shown in Tables \ref{table:process-metrics-description} and  \ref{table:process-extended-metrics-description} on \ref{sec:AppendixA2}, were obtained using the discovery method described in \ref{sec:process-metrics}, and by running queries into the events database as presented in Figure \ref{chapter3-0-2.pdf}.

\subsubsection{Data Partitioning}
\label{sec:cluster-selection}
We used the \textbf{k-means} clustering algorithm to compute new variables based on the partition of the teams across different levels (clusters) of Process Cyclomatic Complexity \textit{(PCC)} and McCabe Cyclomatic Complexity variance \textit{(\textDelta VG)} . 
The decision of how many clusters to use \textbf{(k)} was supported by a detailed analysis of the \textbf{Elbow} and \textbf{Silhouette} methods:

\begin{itemize}

    \item \textbf{Elbow Method.} It is frequently used to optimize the number of clusters in a data set. This heuristic, consists of rendering the explained variation as a function of the number of clusters, and picking the elbow of the curve as the optimal number of clusters to use. In cluster analysis, the elbow method runs \textbf{k-means} clustering on the dataset for a range of values for \textbf{k} (say from 2-10), and then, for each value of \textbf{k} computes an average score for all clusters. The distortion score is computed as the sum of square distances from each point to its assigned center \cite{Purnima2014EBK-Means:WSN}.\\

    \item \textbf{Silhouette Method.} It is a commonly used approach of interpretation and validation of consistency within clusters of data. Provides a concise graphical representation of how well each object has been classified within the corresponding cluster. The Silhouette value is a measure of how similar an object is to its own cluster (cohesion) compared to other clusters (separation). The silhouette can be calculated with any distance metric, such as the Euclidean distance or the Manhattan distance, and ranges from -1 to +1. A high value indicates that the object is well matched to its own cluster and poorly matched to neighboring clusters. The clustering configuration is appropriate if most objects have a high value. If many objects have a low or negative value, then the clustering configuration may have too many or too few clusters and, as such, requires further research before a decision on the optimal number of \textbf{k} clusters is made \cite{Kaoungku2018TheFeatures}.\\

\end{itemize}

\subsubsection{Model Selection with Hyperparameter Optimization}
\label{sec:model-selection}

To build, tune model parameters as recommended \cite{Xia2018HyperparameterEstimation,Minku2017AEstimation}, train, evaluate and select the best-fit classification models presented in Tables \ref{table:model-list1} and \ref{table:model-list2}, we used \textbf{Weka} and the \textbf{Auto-Weka} plugin. \textbf{Weka} (Waikato Environment for Knowledge Analysis) is a popular suite of machine learning software written in Java. It's workbench contains a collection of visualization tools and
algorithms for data analysis and predictive modeling, together with graphical user interfaces for easy access to
this functionality \cite{Jagtap2013CensusWEKA}. \textbf{Auto-Weka } is a plugin that installs as a \textbf{Weka} package and uses Bayesian optimization to automatically instantiate a highly optimized parametric machine learning framework with minimum user intervention \cite{Thornton2017Auto-WEKAWEKA}.

\subsubsection{Model Evaluation}
Several evaluation metrics can be used to assess model quality in terms of false positives/negatives (FP/FN), and true classifications (TP/TN). However, commonly used measures, such as \textbf{Accuracy, Precision, Recall and F-Measure}, do not perform very well in case of an imbalanced dataset or they require the use of a minimum probability threshold to provide a definitive answer for predictions. For these reasons, we used the \textbf{ROC}\footnote{Receiver operating characteristic (\textbf{ROC}) is a curve that plots the true positive rates against the false positive rates for all possible thresholds between 0 and 1.}, which is a threshold invariant measurement. Nevertheless, for general convenience, we kept present in Tables \ref{table:model-list1} and \ref{table:model-list2} all the evaluation metrics.

\textbf{ROC} gives us a 2-D curve, which passes through (0, 0) and (1, 1). The best possible model would have the curve close to y = 1, with and area under the curve (\textbf{AUC}) close to 1.0. \textbf{AUC} always yields an area of 0.5 under random-guessing. This enables comparing a given model against random prediction, without worrying about arbitrary thresholds, or the proportion of subjects on each class to predict \cite{Rahman2013HowBetter}.

\subsection{Research Questions}

The research questions for this work are:
\begin{itemize}
\item \textbf{RQ1:} How different refactoring methods perform when the goal is to reduce complexity, future testing and maintainability efforts?.\\
\textbf{Methods Used.} Process Mining Model Discovery, Descriptive statistics and Cluster Analysis. \\

\item \textbf{RQ2:} Is there any association between software complexity and the underlying development activities in refactoring practices?\\
\textbf{Methods Used.} Process Mining Model Discovery, Correlation Analysis using the Spearman's rank correlation.\\

\item \textbf{RQ3:} Using only process metrics, are we able to predict with high accuracy different refactoring methods?\\
\textbf{Methods Used.} Supervised and Unsupervised Learning Algorithms with Hyperparameter Optimization.\\

\item \textbf{RQ4:} Using only process metrics, are we able to model accurately the expected level of complexity variance after a refactoring task?\\
\textbf{Methods Used.} Supervised and Unsupervised Learning Algorithms with Hyperparameter Optimization.\\


\end{itemize}


\section{Study Results}
\label{sec:study-results}
In this section, we present the experiment results with respect to our research questions.

\subsection{\textbf{RQ1. How different refactoring methods perform when the goal is to reduce complexity, future testing and maintainability efforts?}}


 



In this \textbf{RQ}, we used as product metrics, the ones identified in section \ref{sec:product-metrics}. Since IDE usage is a sequence of actions (it can be seen as a process, or at least, as a process fragment), we used as process metrics the ones identified in \ref{sec:process-metrics}. Notice that both, product and process metrics, have been computed to obtain the \textbf{\textDelta}\ between \textbf{\textbf{t1 and t0}}.

\begin{table}[H]
\footnotesize
\caption{Teams' Statistics}
\label{table:teams-statistics}
\begin{tabular}{lcccccc}
	\hline\noalign{\smallskip}
	\textbf{Task Mode} & \textbf{Teams} & \textbf{Dev.} & \textbf{Ses.} & \textbf{Evts.} &  \textbf{\textDelta VG} & \textbf{PCC}  \\
	\noalign{\smallskip}\hline\noalign{\smallskip}

\textbf{Automatic Refactoring} & 32 & 65 & 150 & 10443  & 7.81\% & 166.5\\[0.2cm]
 \textbf{Manual Refactoring} & 39 & 52 & 170  & 22676 & 2.69\% & 300.3\\[0.2cm]
\hline\\
 \textbf{Total} & 71 & 117 & 320 & 33119 & &\\
   \noalign{\smallskip}\hline
\multicolumn{7}{l}{\shortstack[l]{\textbf{Dev} - Developers, \textbf{Ses} - Sessions, \textbf{Evts} - Events, \\\textbf{\textDelta VG} - McCabe Cyclomatic Complexity Reduction \%(mean),\\ \textbf{PCC} - Process Cyclomatic Complexity(mean)}}
\end{tabular}
\end{table}

\begin{table}[H]
\footnotesize
\caption{Teams' Refactoring Results}
\label{table:process-quantiles}
\begin{tabular}{lcccccc}
	\hline\noalign{\smallskip}
    \textbf{Metric Name} & \textbf{Min.} & \textbf{1st Qu.} & \textbf{Median} & \textbf{Mean} & \textbf{3rd Qu.} &\textbf{Max.}\\
	\noalign{\smallskip}\hline\noalign{\smallskip}
	\multicolumn{7}{l}{\textbf{Automatic Refactoring}}\\[0.2cm]
    \textbf{\textDelta VG} & 2.68\% & 5.87\% & 6.95\% & 7.81\% & 8.84\% & 16.77\%\\[0.2cm]
    \textbf{PCC} & 24.0 & 77.0 & 168.5 & 166.5 & 218.2 & 407.0\\[0.5cm]
   
     \multicolumn{7}{l}{\textbf{Manual Refactoring}}\\[0.2cm]
     \textbf{\textDelta VG} & 0.32\%  &  0.62\% &   0.94\% &   2.69\%  &  3.92\%  & 13.98\%\\[0.2cm]
     \textbf{PCC} & 53.0  & 152.0  & 275.0 &  300.3  & 407.0  & 738.0\\[0.5cm]
 
  \multicolumn{7}{l}{\textbf{Data Partition}}\\[0.2cm]
  \multicolumn{7}{l}{\textbf{VG\_LEVEL} \hspace{0.4cm} \textbf{LOW} = [0\%, 4\%]; \textbf{MEDIUM} = [4.1\%, 9\%]; \textbf{HIGH} = [$>$9\%]}\\[0.2cm]
  \multicolumn{7}{l}{\textbf{PCC\_LEVEL} \hspace{0.15cm} \textbf{LOW} = [0, 285]; \textbf{HIGH} = [$>$285]}\\[0.2cm]
  
   \noalign{\smallskip}\hline
   \multicolumn{7}{l}{\shortstack[l]{\textbf{\textDelta VG} - McCabe Cyclomatic Complexity Reduction \%,\\ \textbf{PCC} - Process Cyclomatic Complexity}}
\end{tabular}
\end{table}

We had 32 teams performing automatic refactoring using the \textit{JDeodorant} plugin, and 39 doing manual refactoring supported only by the Eclipse native features and/or driven by the developers experience and skills. Table \ref{table:teams-statistics} shows the total number of developers and their activities, here referred as development sessions.
In Table \ref{table:process-quantiles} we show measures of central tendency and measures of variability regarding the distribution of \textit{\textDelta VG} and \textit{PCC}, together with how both were partitioned.

Figure \ref{chapter5-4.pdf} provides evidence for selecting the optimal number of clusters to partition the data according to \textbf{LOW} or \textbf{HIGH} levels of process cyclomatic complexity used in Figure \ref{chapter4-2.pdf}. The same clustering method was used to partition the different levels of software cyclomatic complexity as \textbf{LOW}, \textbf{MEDIUM} or \textbf{HIGH}.\\

\plot{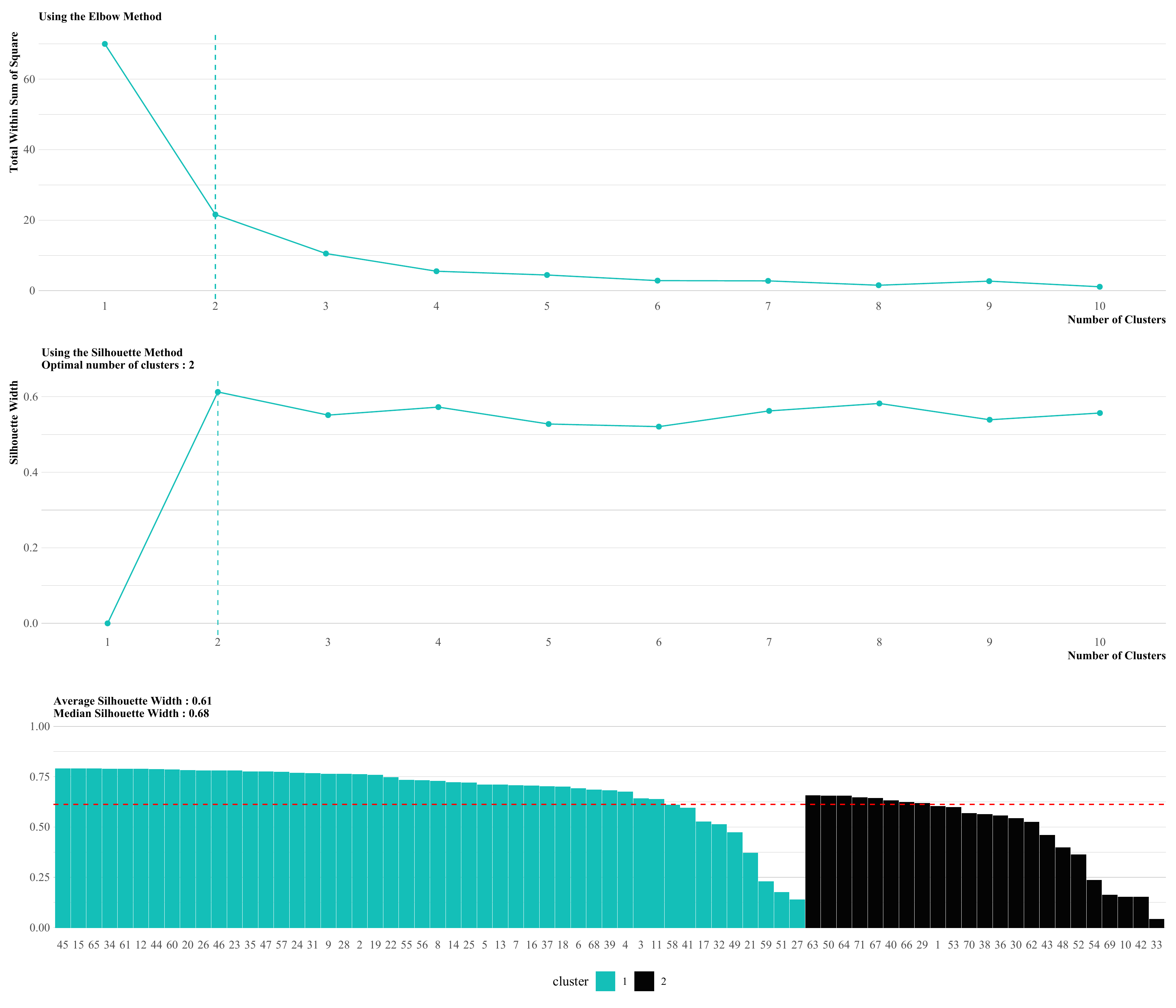}[Detecting optimal partitions of PCC][13.5][ht!][trim=0.2cm 0cm 0.2cm 0.2cm]

\plot{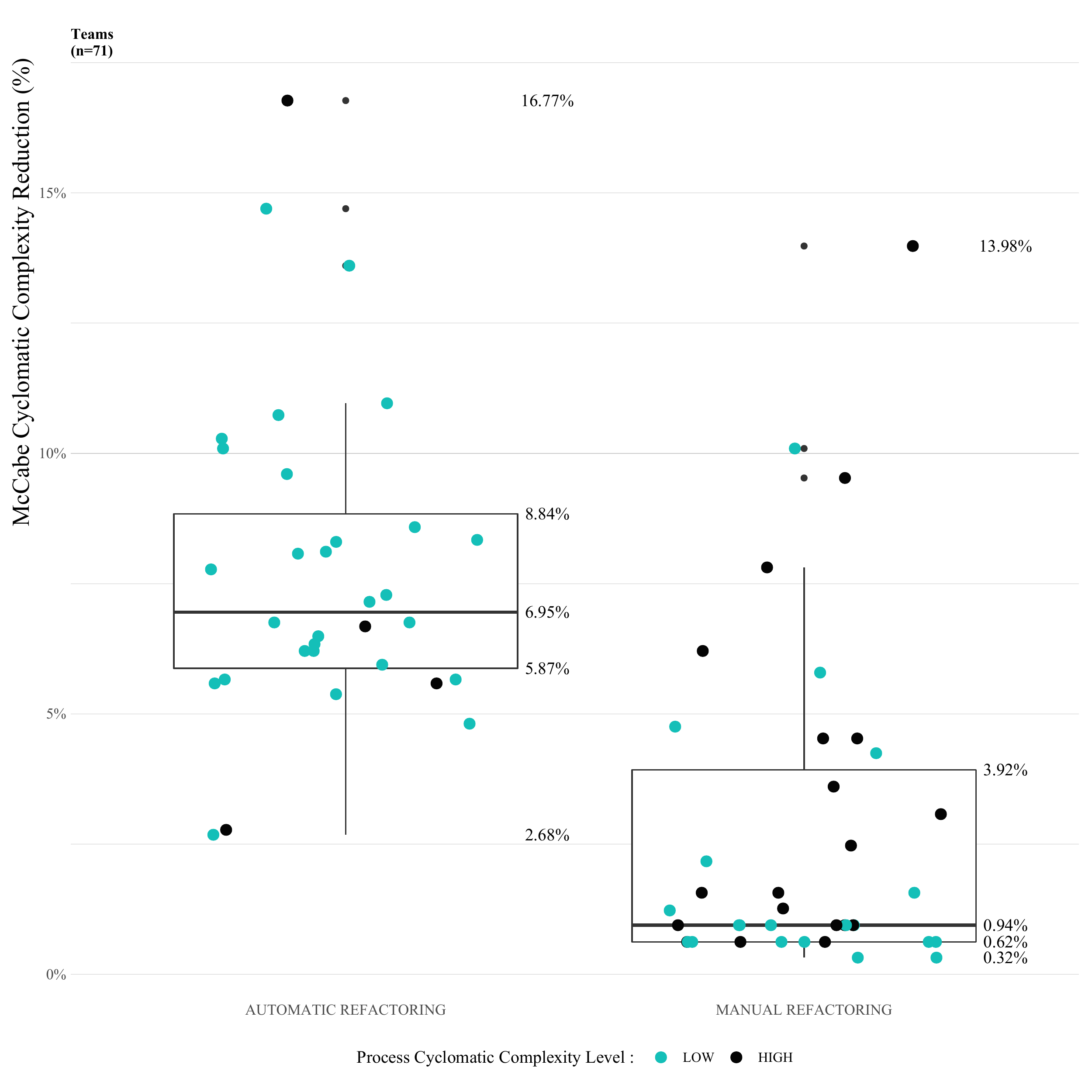}[Refactoring Practices Comparison][13][ht!][trim=0.2cm 0cm 1cm 0.3cm]

\plot{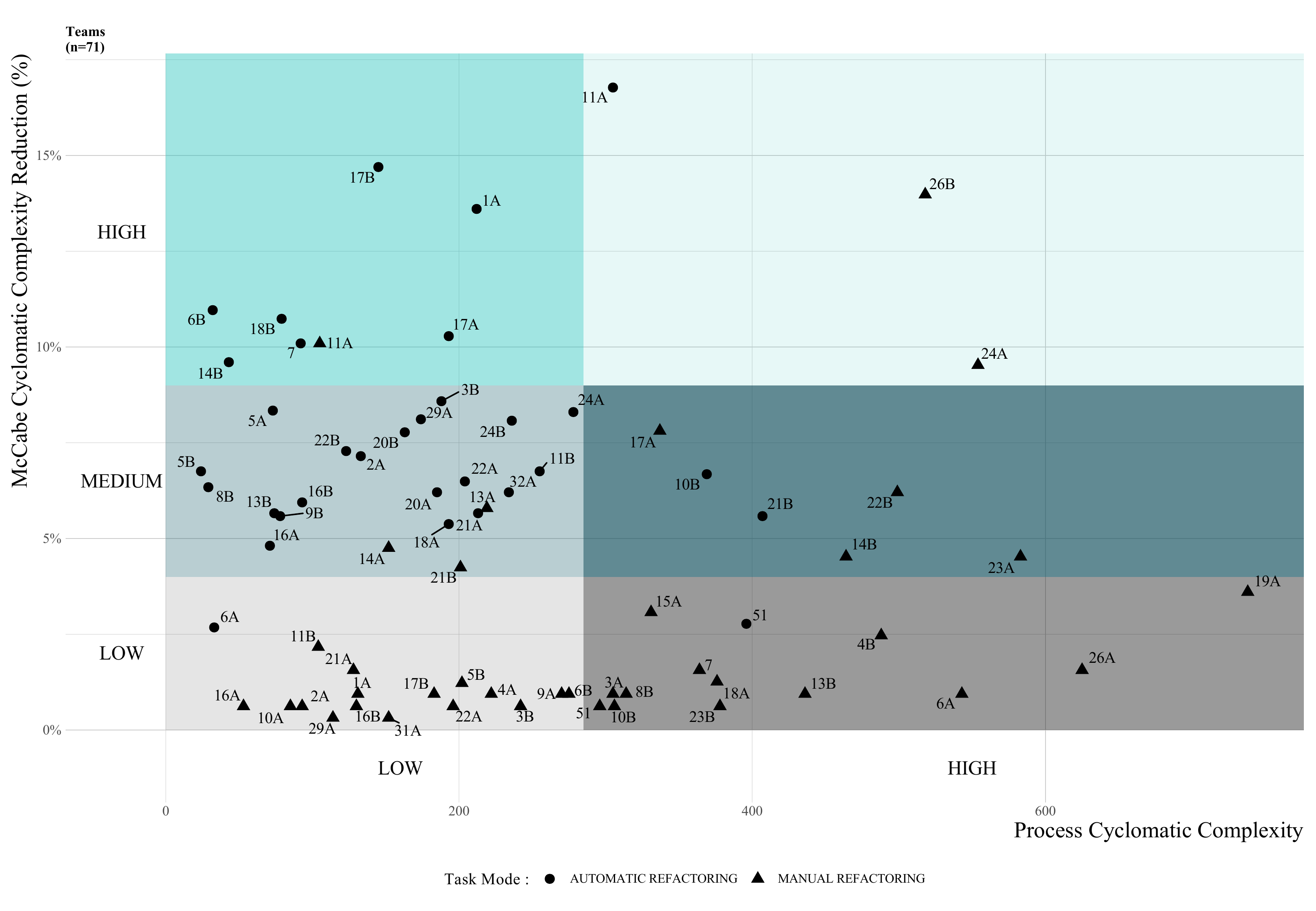}[Plotting teams according to levels of software and process cyclomatic complexity][13.5][ht!][trim=0cm 0.3cm 0cm 0.4cm]

\begin{figure}
  \centering   
  \begin{overpic}[width=13.8cm]{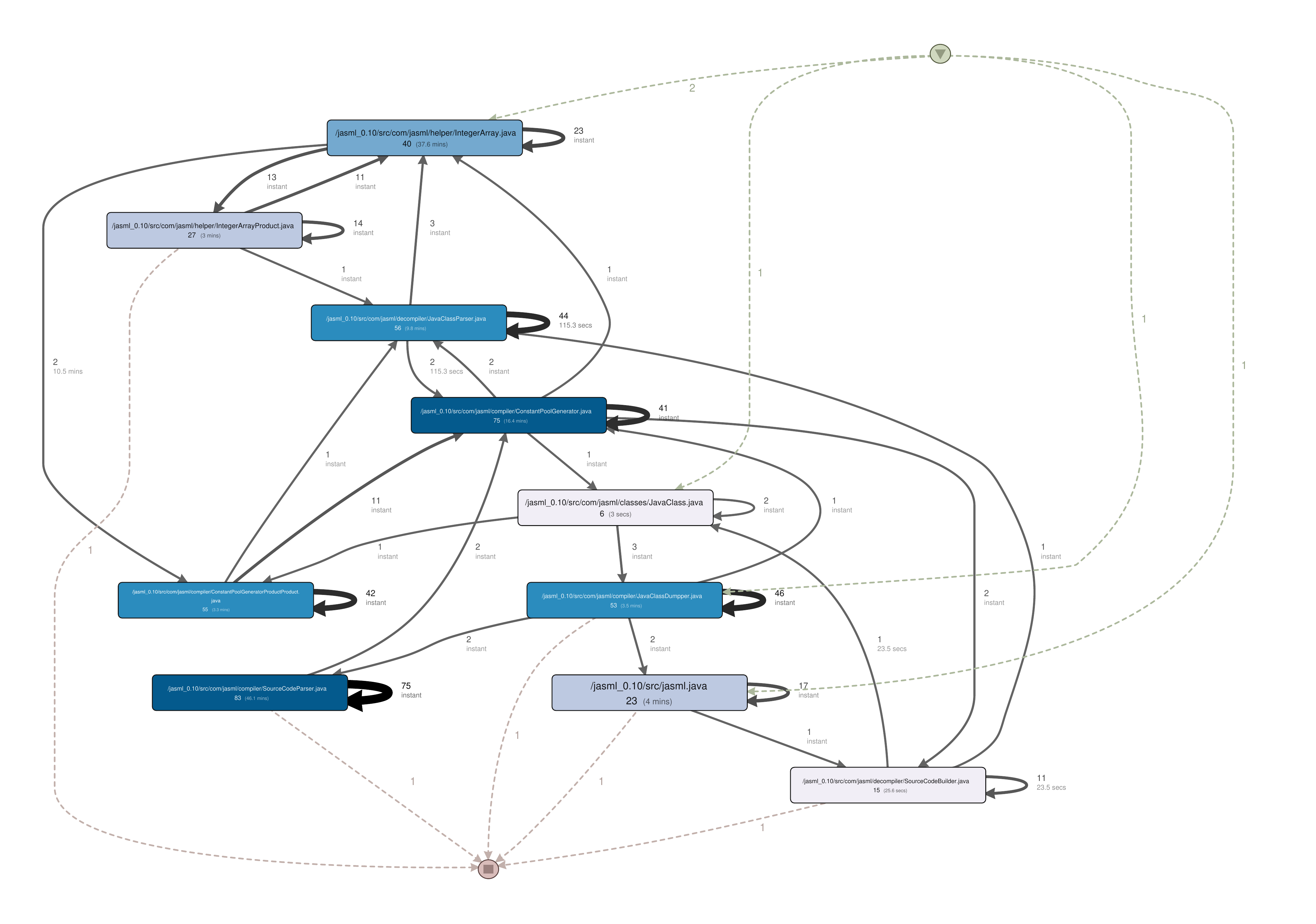}
     \put(5,65){\includegraphics[width=3.5cm, clip, trim=2.1cm 0.5cm 1cm 2cm]{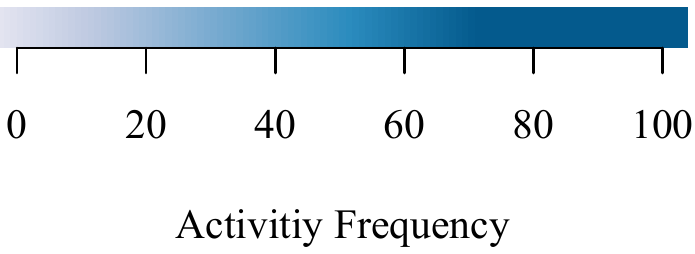}}
     \put(33,65){\includegraphics[width=3.5cm, clip, trim=2.1cm 0.5cm 1cm 2cm]{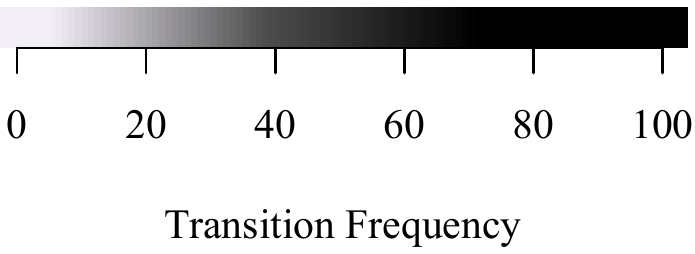}}  
  \end{overpic}
\caption{Team 11A : High PCC and High VG reduction (20\% activities/files, 80\% paths)}
\label{chapter4-10-Flow.pdf}
\end{figure}

\newpage
\textbf{Observation 1: Automatic Refactoring achieves higher levels of McCabe Cyclomatic complexity reduction.} Consider relevant in Table \ref{table:teams-statistics}, how the mean of code cyclomatic complexity reduction \textit{(\textDelta VG)} for automatic refactoring is almost three times the reduction when doing manual refactoring. It is also relevant to mention, by looking at Figure \ref{chapter4-2.pdf}, that only four teams had high complexity levels in their work sessions when doing refactoring using \textit{JDeodorant}. Furthermore, from those, one team had the major software complexity reduction(16.77\%), whilst other had near the lowest value of reduction(2.68\%) within the automatic refactoring practice.
The observation of such different results raised the doubt about the comprehension, focus and behaviour of those two teams in the given task. This demanded further investigation on their efficiency, for which, we provide some evidences later using Figures \ref{chapter4-10-Flow.pdf} and \ref{chapter4-11-Flow.pdf}.\\

\textbf{Observation 2: Manual refactoring practices have higher process cyclomatic complexity.} We observe that teams doing manual refactoring almost double the mean of process cyclomatic complexity \textit{(PCC)}, when compared with the ones using the automatic features of \textit{JDeodorant}. Being deprived of the code smell detection plugin, these teams had to do more manual work to potentially achieve the same results as the ones doing automatic refactoring. This suggest that the refactoring plugin was working as expected, thus reducing software complexity with less effort simply because several code snippets may have been introduced automatically.

On the contrary, teams doing the task manually needed to do more code, and therefore, more actions within the IDE to detect and correct the code smells. As shown earlier in section \ref{sec:introduction}, manual refactoring tasks can introduce non expected defects in the code and is seen as a practice to avoid.


Figure \ref{chapter4-2.pdf} plot the percentage of McCabe Cyclomatic Complexity per method reduction obtained after both refactoring sessions. 
The different colors plot the different levels of process cyclomatic complexity as discovered from mining each team events log.\\

\plot{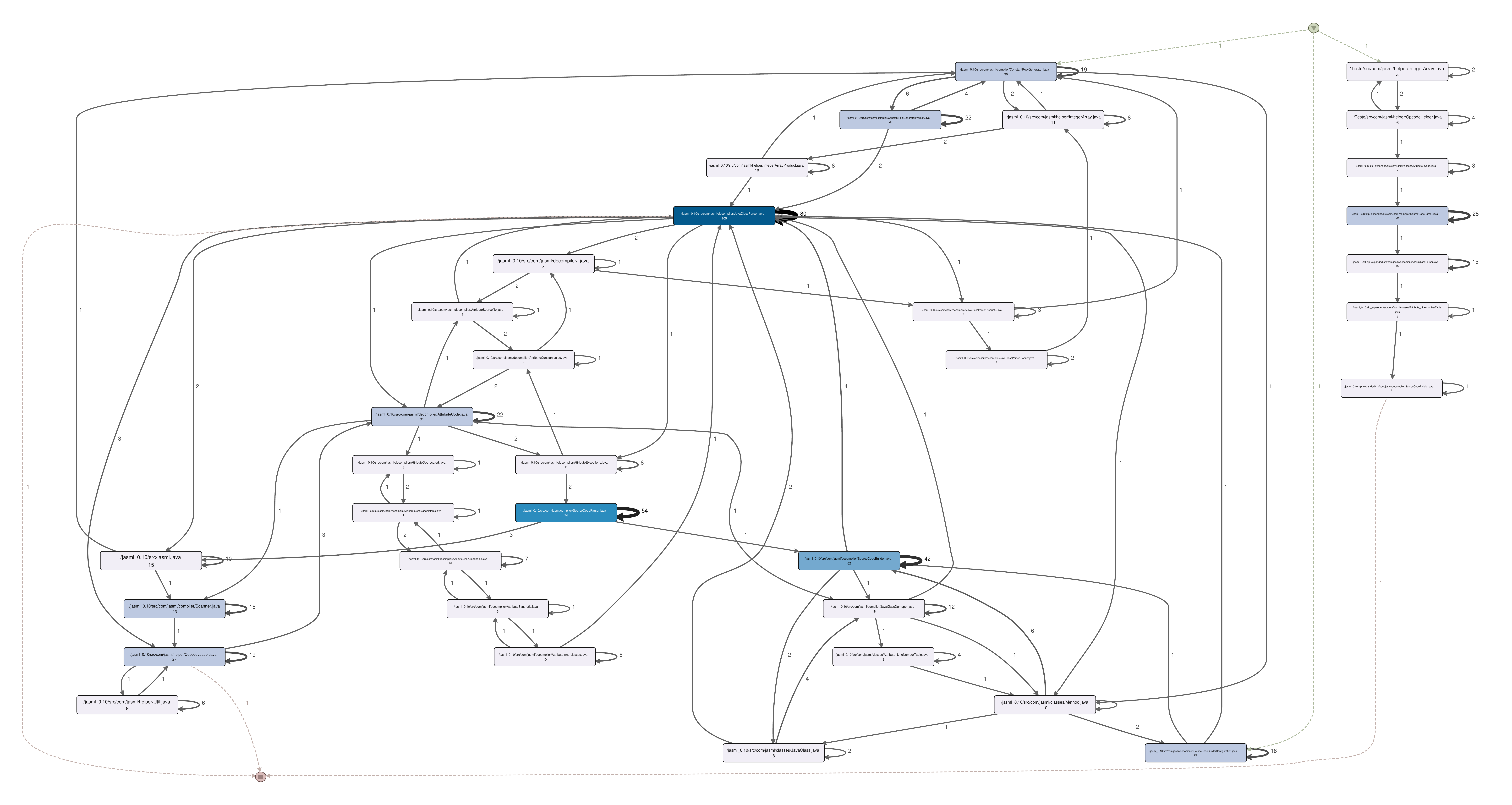}[Team 51 : High PCC but Low VG reduction (20\% activities/files, 80\% paths)][13.5][ht!][trim=1.3cm 1.3cm 2.5cm 1.3cm]

\newpage
\textbf{Observation 3: Even using JDeodorant, similar work efforts does not mean the same level of gains in software complexity reduction.} If it is apparent that, when using \textit{JDeodorant}, the processes tend to have lower levels of complexity and obtained globally more gains in product complexity reductions, the same cannot be said for teams doing manual refactoring. These teams have a more heterogeneous process behavior since they were free to apply refactoring functionalities without any guidelines in detection and correction from a dedicated plugin. Figure \ref{chapter4-c-1.pdf} identifies all teams and distributes them according to their levels of software and process complexity.


From Figure \ref{chapter4-2.pdf}, we can also observe that the team (11A) with the highest reduction in code complexity ($\approx$ 16.77\%), had also a high level of process complexity even if they were using the JDeodorant plugin. We can also identify a team(51) doing automatic refactoring with high levels of process complexity but having instead, very low gains in code cyclomatic complexity reduction ($\approx$ 2.68\%). As such we investigated the activities of both teams in order to identify potential reasons for this substantial variation.

Figures \ref{chapter4-10-Flow.pdf} and \ref{chapter4-11-Flow.pdf}, represent the process flow views of both individual teams regarding the files browsed and/or changed during the refactoring practice\footnote{We acknowledge that the labels in these two diagrams, produced by the Disco tool, are illegible in a printing version. However, since the figures are in vectorial format, they can be \textit{zoomed in} easily if this paper is read in its electronic version (pdf), the most probable access medium.}. Based on the same values for the activities and paths, we can clearly identify that the team with high gains in \textit{VG} reduction worked in less files (number of nodes) and was focused evenly on all of them (dark blue nodes means more actions on those files).

\plot{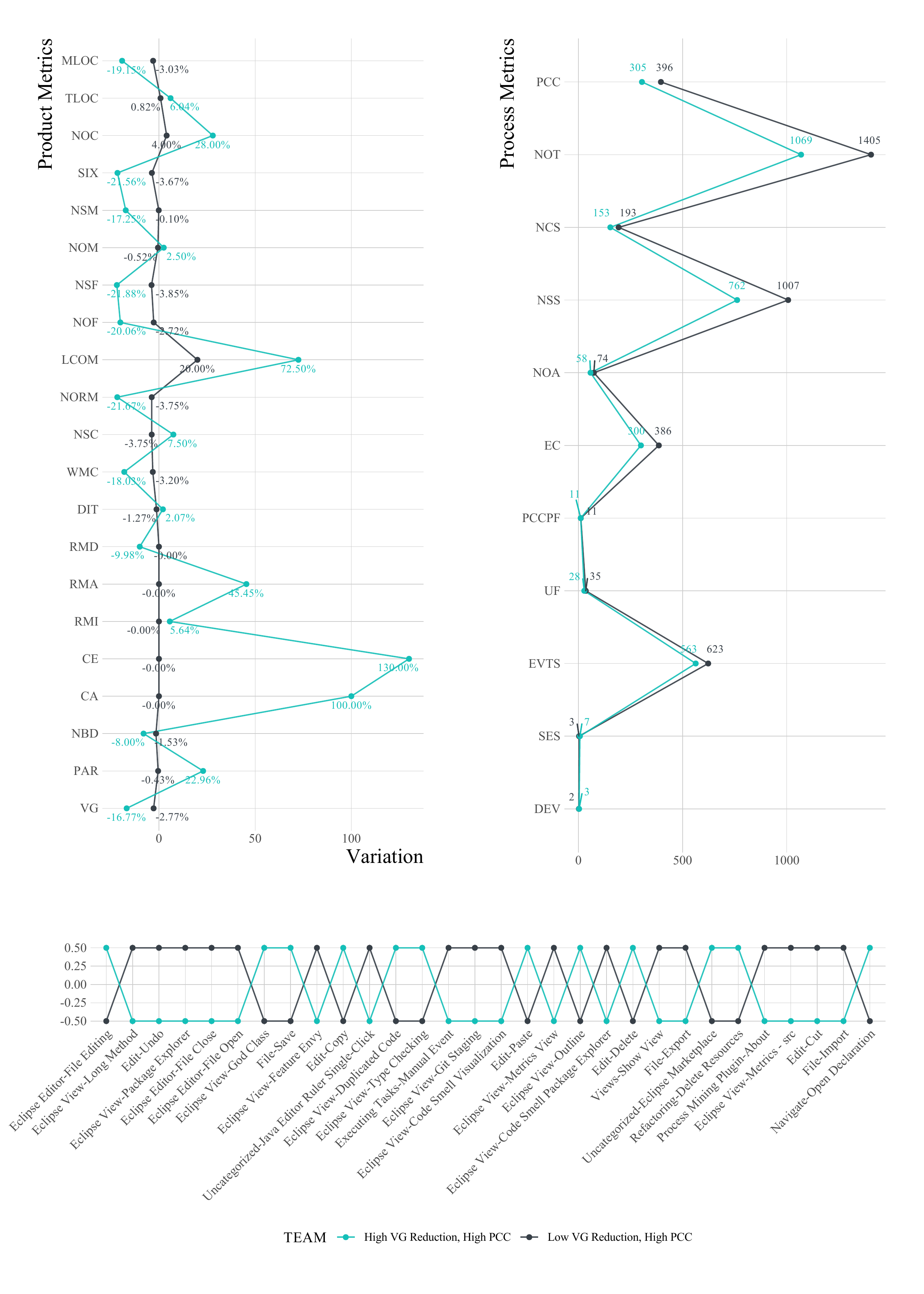}[Teams(11A vs. 51) with distinct VG variance positioning but similar PCC levels][11][ht!][trim=0.2cm 1cm 1cm 1cm]

On the contrary, the team with low gains in \textit{VG}, visited more files but worked frequently on only 3 of them. This fuzzy behavior suggests lack of focus and/or knowledge about the task to accomplish, and present a good way to measure efficiency on development teams or individual developers.
That can be confirmed by comparing both teams statistics in Figure \ref{chapter4-111.pdf}, where we present product metrics, process metrics and extended process metrics scaled to represent their position to the mean value of each action for both teams.

We highlight in the extended process metrics the fact that the team with bigger VG reduction was the one with less frequencies in commands such as : Undo, Cut, File Open, File Close plus other navigational and less productive actions.
This team had also bigger frequencies in commands to detect and fix code smells, such as: God Class, Duplicated Code and Type Checking.
However, the gains in the \textit{VG} reduction were achieved at the cost of increasing  28\% the number of classes(NOC) and the lack of cohesion of methods(\textit{LCOM}) by $\approx$72\%. On the process side, despite the fact that this team had more work sessions(7), they touched less files, meaning their activities were less complex, and that is confirmed by the \textit{UF}, \textit{NOA} and \textit{PCC} metrics.



\subsection{\textbf{RQ2. Is there any association between software complexity and the underlying development activities in refactoring practices?}}
With the evidences shown in \textbf{RQ1} for the two distinct refactoring methods, one may question if the product complexity reduction gains are monotonically correlated with the development activities which originated them.


We used the Spearman correlation coefficient to measure the strength of correlation between metrics of these two dimensions, product and process complexities. This coefficient ranges from -1 to 1, where -1 and 1 correspond to perfect negative and positive relationships respectively, and 0 means that the variables are independent of each other.


To validate our results, we performed a significance test to decide whether based upon this sample there is any or no evidence to suggest that linear correlation is present in the population. As such, we tested the null hypothesis, \hypnull, and the alternative hypothesis, \hypalt, to gather indication of which of these opposing hypotheses was most likely to be true.

Let \textbf{p\textsubscript{s}} be the Spearmans' population correlation coefficient both for automatic and manual refactoring, then we can thus express this test as:

\indent\hypnull: p\textsubscript{s} = 0 : No monotonic correlation is present in the practice.\\
\indent\hypalt: p\textsubscript{s} $\neq$ 0: A monotonic correlation is present in the practice.\\

\noindent\textbf{Automatic Refactoring.} After computing the Spearman correlation coefficient on the subset of teams doing automatic refactoring, and despite the fact that some correlations were slightly negative as we expected, we got no significant statistics on the correlation of \textit{\textDelta VG} and \textit{PCC} or any other pair of metrics, as shown by Spearmans' \textit{rho} and \textit{p-value} in Table \ref{table:spearman-both-correlations}.\\ 

\textbf{Observation 4: No significant correlation was found between product and process metrics on automatic refactoring practices.}
Hence, we can say that we cannot reject the null hypothesis, \hypnull, meaning that a monotonic correlation cannot said to be found between code cyclomatic complexity and process cyclomatic complexity or any other process metric.

\begin{table}[H]
\footnotesize
\caption{Spearmans' Correlation - Refactoring Tasks}
\label{table:spearman-both-correlations}
\begin{tabular}{lcccc}
	\hline\noalign{\smallskip}
	& \multicolumn{2}{c}{\cellcolor{gray!10}\textbf{Automatic Refactoring}} &   \multicolumn{2}{c}{\cellcolor{gray!10}\textbf{Manual Refactoring}}\\
	& \multicolumn{2}{c}{\cellcolor{gray!10}\textbf{\textDelta VG}} &  \multicolumn{2}{c}{\cellcolor{gray!10}\textbf{\textDelta VG}}\\
	\textbf{Factors} & \textbf{\textit{Spearmans' rho}} & \textbf{\textit{p-value}} &  \textbf{\textit{Spearmans' rho}} & \textbf{\textit{p-value}}\\
	\noalign{\smallskip}\hline\noalign{\smallskip}

 \textbf{PCC} & -0.02 & 0.9707 & \textbf{0.43} & \textbf{0.0432*}\\[0.2cm]
\textbf{UF} & 0.01 & 0.5218 &  0.32 & 0.3427\\[0.2cm]
\textbf{SES} & 0.15 & 0.7489 &  0.24 & 0.2814\\[0.2cm]
\textbf{DEV} & -0.05 & 0.7342 &  0.03 & 0.8193\\[0.2cm]
\textbf{NPER} & -0.19 & 0.4976 &  \textbf{0.32} & \textbf{0.0197*}\\[0.2cm]
\textbf{NISP} & -0.10 & 0.6875 &  \textbf{0.35} & \textbf{0.0120*}\\[0.2cm]
\textbf{PCCPF} & -0.01 & 0.7787 &  \textbf{0.45} & \textbf{0.0059*}\\[0.2cm]
\textbf{NCAT} & -0.11 & 0.6309 &  \textbf{0.39} & \textbf{0.0096*}\\[0.2cm]
\textbf{NCOM} & -0.05 & 0.6240 &  0.42 & 0.0712\\[0.2cm]


   \noalign{\smallskip}\hline
   	\multicolumn{3}{l}{\textit{*Statistically significant if p-value $<$ 0.05}}
\end{tabular}
\end{table}


\noindent\textbf{Manual Refactoring.} When analyzing the dataset with manual refactoring activities, we found that product complexity reduction was moderately correlated with the process cyclomatic complexity and several other metrics process related. Table \ref{table:spearman-both-correlations} presents Spearmans' \textit{rho} and \textit{p-value}, highlighting the significant correlations\footnote{Other product and process metrics were omitted due to the absence of significant correlations}.\\

\textbf{Observation 5: A moderate correlation was found between product metrics and process metrics on manual refactoring tasks.}
It is relevant to highlight the presence of a moderate positive correlation between the product cyclomatic complexity reduction \textit{(\textDelta VG)} and the overall process cyclomatic complexity\textit{(PCC)} and per unique file touched\textit{(PCCPF)}. This means that the more actions the teams have done within the IDE the bigger the gains obtained in complexity reduction. 

\plot{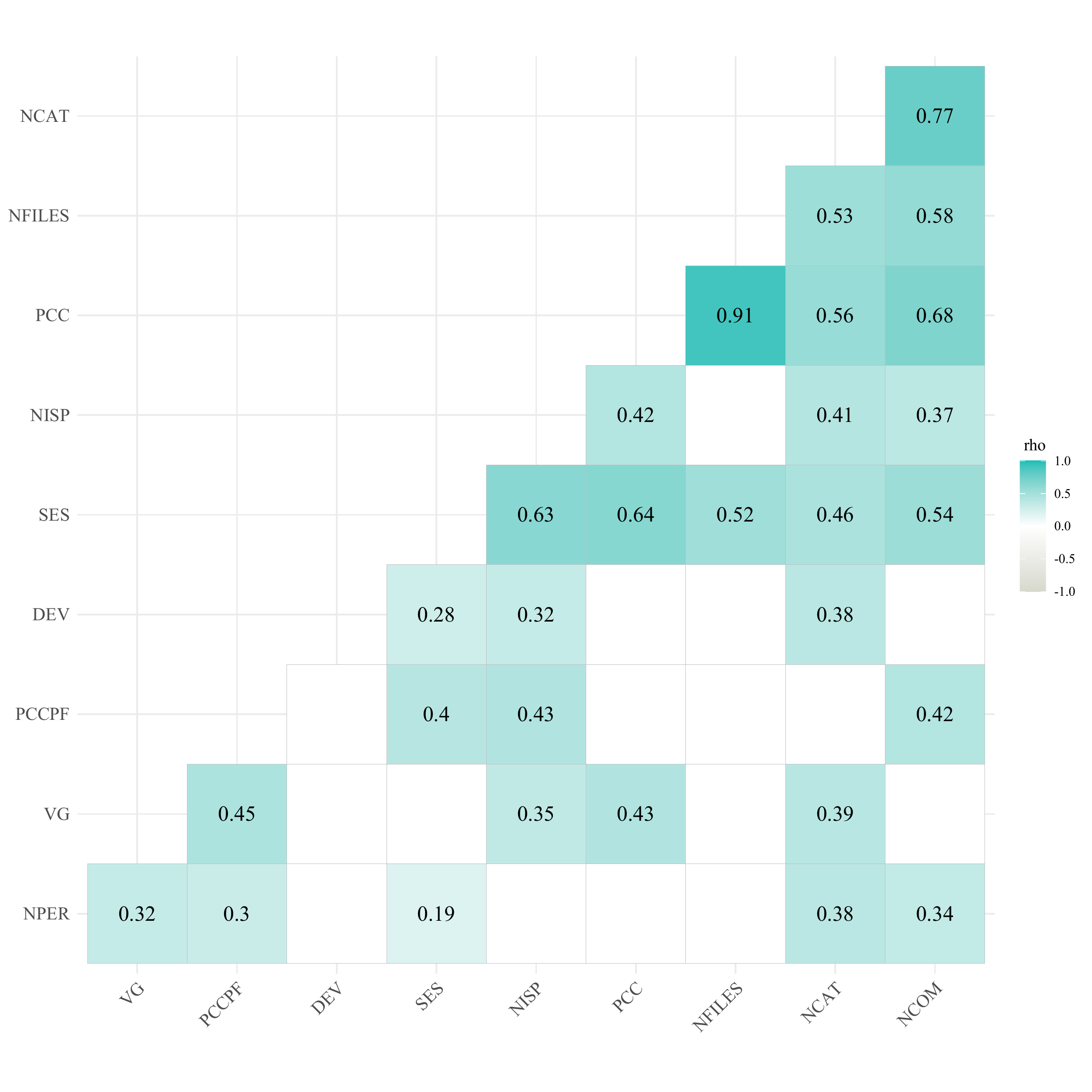}[Manual Refactoring correlation results][13.5][H][trim=0cm 0cm 0cm 0cm]

\textbf{Observation 6: Weak to moderate correlations were found between product complexity reduction and IDE command categories.}
Weak to moderate correlations emerge when we pair the product complexity reduction with the number of the IDE command categories\textit{(NCAT)}, IDE perspectives activated\textit{(NPER)} and the number of distinct physical locations from where the task was performed\textit{(NISP)}. Based on the significance tests, we can reject \hypnull, and accept \hypalt, meaning that a monotonic correlation exists between code cyclomatic complexity and process cyclomatic complexity as well as with the other highlighted metrics.

\newpage

\textbf{Observation 7: No significant correlations were found between any process metrics and product metrics, except for \textDelta VG.}
All product and process metrics collected are shown in Tables \ref{table:product-metrics-description} and \ref{table:process-metrics-description}.

Figure \ref{chapter4-1.pdf} plot only the significant correlations\footnote{Blank squares means non significant values} among all those we studied. As expected, process metrics show strong correlations between themselves, however, we find this result obvious and not relevant withing the context of this study. 

\subsection{\textbf{RQ3. Using only process metrics, are we able to predict with high accuracy different refactoring methods?.}}

Process metrics have been confirmed as suitable predictors for many software development prediction models. They were found not only suitable, they performed significantly better than code metrics across all learning techniques in several studies \cite{Madeyski2015WhichStudy,Rahman2013HowBetter}.

Our goal was to use the process metrics described in Table \ref{table:process-metrics-description}, to predict if a refactoring task executed by a group of teams had been done automatically, using the \textit{JDeodorant} features, or manually, using only the Eclipse native functionalities or driven by developers skills. Each subject in our dataset has the class to predict labelled as \textbf{AR} and \textbf{MR} for automatic and manual refactoring, respectively. In this case, we did not use metrics from Table \ref{table:process-extended-metrics-description} because that would introduce bias in our models since the process extended metrics can easily be used to understand if developers used or not IDE built in features or their own skills during a refactoring practice.

Table \ref{table:model-list1} present the results for the 5 best models we got out of the $\approx$30,000 we evaluated on our research. 
In this context, the machine learning models used were built by assembling and testing supervised or unsupervised algorithms adjusted with feature selection and hyperparameter optimization. From the models built, the ones with higher ROC were chosen.
A brief explanation of each algorithm can be found in \ref{sec:AppendixA3}, as well as the code obtained from training Model 1.\\

\textbf{Observation 8: Random Forest confirms its accuracy.} Random Forest models were found to be the ones with higher accuracy in predicting refactoring opportunities in previous studies \cite{Aniche2020TheRefactoring}. We observe the same behaviour. Random Forest shows twice in the top 5 of our best models, with a \textbf{ROC} value of 0.983 and 0.939 for Model 1 and 2, respectively. In both cases, the models were computed by a meta learner which builds an ensemble of randomizable base classifiers, the Random Committee. 

\begin{table}[H]
\footnotesize
\caption{Detailed Model Evaluation}
\label{table:model-list1}
\begin{tabular}{lcccccccc}
	\hline\noalign{\smallskip}
			\textbf{Model} & \textbf{TP} & \textbf{FP} & \textbf{Pre.} & \textbf{Rec.} & \textbf{F-M.} & \textbf{MCC} & \textbf{ROC} & \textbf{PRC}\\
	\noalign{\smallskip}\hline\noalign{\smallskip}
	
	\multicolumn{9}{l}{\cellcolor{gray!10}\textbf{Model 1, RandomCommittee/RandomForest, Accuracy = 92.95\%}}\\
	\textbf{AR} & 0.906  &  0.051  &  0.935   &   0.906  &  0.921   &   0.858 &  0.983  &   0.980 \\[0.1cm]
\textbf{MR} & 0.949  &  0.094  &  0.925   &   0.949 &   0.937  &    0.858  &  0.983   &  0.987 \\[0.1cm]
\textbf{W. Avg.} & 0.930  &  0.075  &  0.930   &   0.930  &  0.929  &    0.858  &  0.983  &   0.984  \\[0.2cm]
	
	\multicolumn{9}{l}{\cellcolor{gray!10}\textbf{Model 2, RandomCommittee/RandomForest, Accuracy = 90.14\%}}\\
	\textbf{AR} & 0.875  &  0.077 &   0.903   &   0.875 &   0.889   &   0.801  &  0.939  &   0.923 \\[0.1cm]
\textbf{MR} & 0.923  &  0.125  &  0.900  &    0.923  &  0.911  &    0.801  &  0.939   &  0.948 \\[0.1cm]
\textbf{W. Avg.} & 0.901 &   0.103  &  0.901   &   0.901  &  0.901   &   0.801  &  0.939  &   0.937  \\[0.2cm]
    
    \multicolumn{9}{l}{\cellcolor{gray!10}\textbf{Model 3, Logistic Model Trees, Accuracy = 90.14\%}}\\
	\textbf{AR} & 0.906  &  0.103  &  0.879   &   0.906  &  0.892   &   0.802  &  0.945  &   0.938 \\[0.1cm]
\textbf{MR} & 0.897  &  0.094  & 0.921   &   0.897  &  0.909    &  0.802  &  0.945 &    0.951  \\[0.1cm]
\textbf{W. Avg.} & 0.901  &  0.098  &  0.902   &   0.901 &   0.902    &  0.802  &  0.945  &   0.945  \\[0.2cm]
	
	\multicolumn{9}{l}{\cellcolor{gray!10}\textbf{Model 4, RandomSubSpace/REPTree, Accuracy = 88.73\%}}\\
	\textbf{AR} & 0.844  &   0.077  &   0.900   &    0.844 &    0.871  &     0.772  &   0.929   &   0.907 \\[0.1cm]
\textbf{MR} & 0.923  &   0.156  &   0.878   &    0.923  &   0.900   &    0.772  &   0.929  &    0.935  \\[0.1cm]
\textbf{W. Avg.} & 0.887  &   0.120  &   0.888  &  0.887  &   0.887  &     0.772  &   0.929  &    0.922  \\[0.2cm]
	
	\multicolumn{9}{l}{\cellcolor{gray!10}\textbf{Model 5, Logistic Regression, Accuracy = 83.09\%}}\\
	\textbf{AR} & 0.750   & 0.103  &  0.857   &   0.750  &  0.800   &   0.659  &  0.939  &   0.940 \\[0.1cm]
\textbf{MR} & 0.897  &  0.250  &  0.814  &    0.897  &  0.854   &   0.659  &  0.939   &  0.950 \\[0.1cm]
\textbf{W. Avg.} & 0.831 &  0.184  &  0.833  &    0.831  &  0.829   &   0.659  &  0.939  &   0.945  \\[0.2cm]

   \noalign{\smallskip}\hline
   \multicolumn{9}{l}{\textbf{TP}-True Positive, \textbf{FP}-False Positive, \textbf{Pre}-Precision, \textbf{Rec}-Recall,}\\
   \multicolumn{9}{l}{  \textbf{F-M}-F-Measure, \textbf{MCC}-Matthews Correlation Coefficient,}\\
   \multicolumn{9}{l}{ \textbf{ROC}-Receiver Operating Characteristic, \textbf{PRC}-Precision-Recall Curve, }\\
   \multicolumn{9}{l}{ \textbf{AR}-Automatic Refactoring, \textbf{MR}-Manual Refactoring,}\\
   \multicolumn{9}{l}{ \textbf{W. Avg}-Weighted Average}
\end{tabular}
\end{table}

Our dataset is not imbalanced, thus, we have almost the same number of subjects for each class, meaning we may use also the \textbf{Accuracy} metric to complement our analysis. Model 1 and 2 had respectively, an accuracy of 92.5\% and 90.14\%.\\



\plot{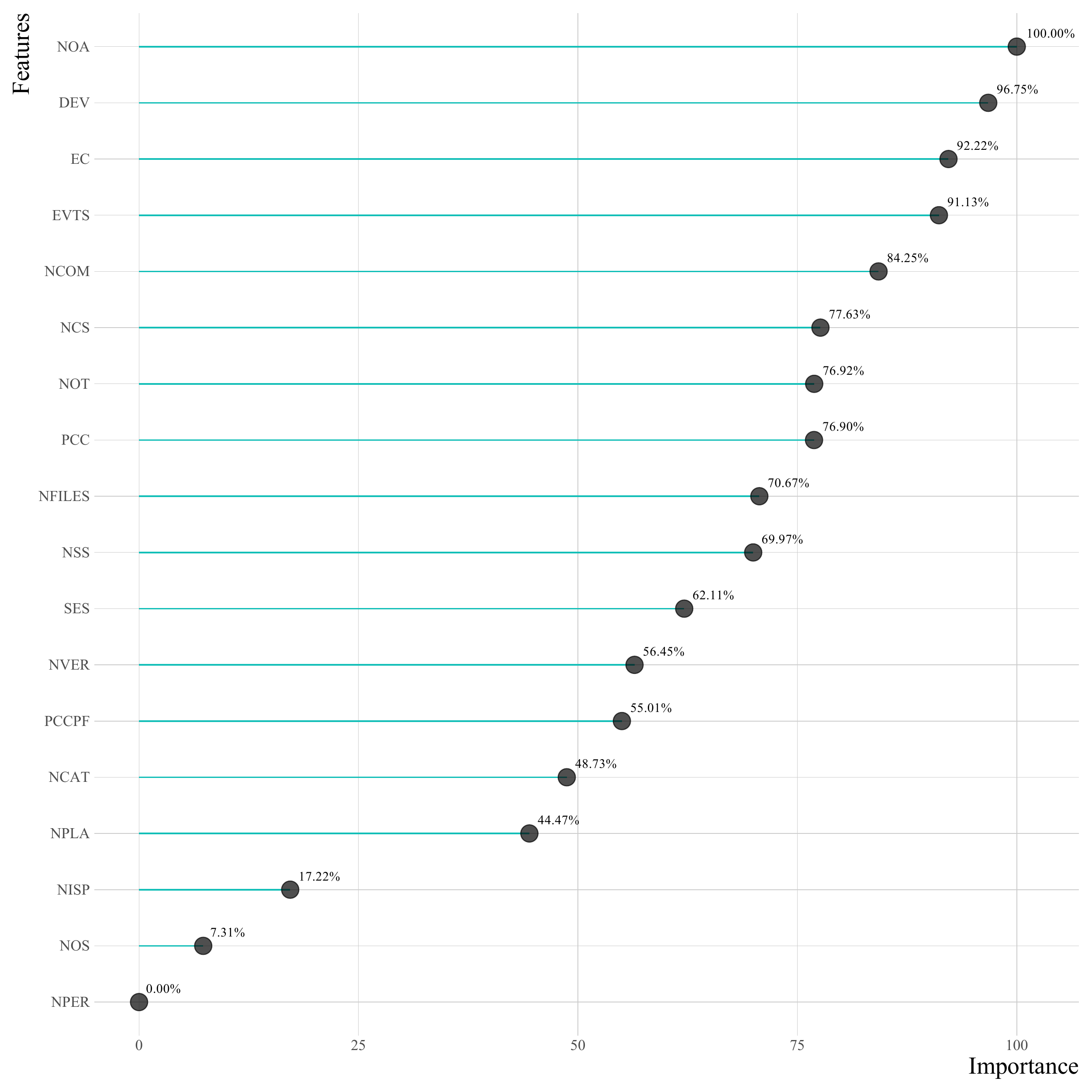}[Feature importance for models on Table \ref{table:model-list1}][13.5][H][trim=0cm 0cm 0cm 0cm]

During models computation phase, we also assessed which of the features were more or less important to predict the refactoring practices: automatic(\textbf{AR}) or manual(\textbf{MR}). Figure \ref{chapter5-b-1.pdf} shows their average importance.

\textbf{Observation 9: Number of Activities, Developers and Commands are the most relevant model features.} These features show among the ones with highest importance in the models we computed. We recall that the number of activities (\textit{NOA}) is a composite metric obtained by the process mining extraction plugin using a hierarchical structure composed of the filename, command category and commands issued during the coding phase. Having a mid level importance we find the process cyclomatic complexity and the number of development sessions.

\textbf{Observation 10: Distinct IDE Perspectives and Operating Systems have almost irrelevant importance.} In our models, the different types operating system used by the developers, the different number of IDE perspectives and number of development locations (\textit{NISP}) are irrelevant predictors in modeling the type of refactoring performed. We argue that, particularly the number of different locations from where the developers performed their work require additional research in order to get any generalized conclusions about this insight.\\






\subsection{\textbf{RQ4: Using only process metrics, are we able to model accurately the expected level of complexity variance after a refactoring task?}}

To answer this \textbf{RQ}, we used not only the metrics from Table \ref{table:process-metrics-description}, but also the ones from Table \ref{table:process-extended-metrics-description}. 
During our analysis, it was clear that process extended metrics, representing the commands issued by each developer/team, added significant predictive power to the models computed. Therefore, to predict the expected software cyclomatic complexity we needed to include individual commands frequencies in addition to the process metrics used in previous \textbf{RQ}. 
By doing this we were able to achieve models with higher accuracy and good ROC values. However, in general, these models have lower accuracy than the ones in \textbf{RQ3}.

Table \ref{table:model-list2} shows the top five models computed to predict the complexity level gains obtained after a refactoring session, either using a dedicated plugin or simply by using Eclipse features.

\textbf{Observation 11: Locally Weighted Learning combined with a Decision Table outperforms Random Forrest.} Contrary to the previous RQ, in this case the best model is not based on a Random Forrest algorithm. However, the latter show as the second best model in terms of accuracy.
The Locally Weighted Learning method uses an instance-based algorithm to assign instance weights which are then used by a specified weighted instances handler. It uses a stack of methods, initially a cluster like mechanism such as the LinearNNSearch and then a Decision Table to classify the outcome. This shows up at no surprise since Decision Tables use the simplest hypothesis spaces possible and usually outperform state-of-the-art classification algorithms.

\begin{table}[H]
\footnotesize
\caption{Detailed Model Evaluation}
\label{table:model-list2}
\begin{tabular}{lcccccccc}
	\hline\noalign{\smallskip}
			\textbf{Model} & \textbf{TP} & \textbf{FP} & \textbf{Pre.} & \textbf{Rec.} & \textbf{F-M.} & \textbf{MCC} & \textbf{ROC} & \textbf{PRC}\\
	\noalign{\smallskip}\hline\noalign{\smallskip}
	
	\multicolumn{9}{l}{\cellcolor{gray!10}\textbf{Model 1, LWL/LinearNNSearch/DecisionTable, Accuracy = 94.36\%}}\\

     \textbf{LOW} & 0.968 &   0.000 &   1.000  &    0.968  &  0.984     & 0.972  &  0.991 &    0.992 \\[0.1cm]
     \textbf{MEDIUM} & 1.000 &    0.095 &   0.879    &  1.000 &   0.935    &  0.892  &  0.994   &  0.992 \\[0.1cm]
     \textbf{HIGH} & 0.727  &  0.000 &    1.000 &     0.727 &   0.842  &    0.832  &  0.992   &  0.967 \\[0.1cm]
     \textbf{Weighted Avg.} & 0.944 &   0.039    & 0.950    &  0.944 &    0.942  &    0.917    & 0.993 &    0.988  \\[0.1cm]

	\multicolumn{9}{l}{\cellcolor{gray!10}\textbf{Model 2, Bagging/RandomForest, Accuracy = 83.09\%}}\\

     \textbf{LOW} & 0.839  &  0.075  &  0.897  &    0.839 &   0.867 &     0.771  &  0.938  &   0.94 \\[0.1cm]
     \textbf{MEDIUM} & 0.828 &    0.095  &  0.857    &  0.828 &   0.842  &    0.737  &  0.971 &     0.945 \\[0.1cm]
     \textbf{HIGH} & 0.818   & 0.083    & 0.643  &    0.818 &   0.720    &  0.668 &   0.971    & 0.827 \\[0.1cm]
     \textbf{Weighted Avg.} & 0.831 &   0.085  &  0.841  &    0.831  &  0.834  &    0.741  &  0.957  &   0.926  \\[0.1cm]

    \multicolumn{9}{l}{\cellcolor{gray!10}\textbf{Model 3, KStar, Accuracy = 78.87\%}}\\

     \textbf{LOW} & 0.935  &  0.225   & 0.763    &  0.935    & 0.841  &    0.707 &   0.948 &    0.951 \\[0.1cm]
     \textbf{MEDIUM} & 0.862    & 0.143 &   0.806 &     0.862 &   0.833  &    0.713  &  0.945 &    0.915 \\[0.1cm]
     \textbf{HIGH} & 0.182   & 0.000  &  1.000    &  0.182   & 0.308    &  0.398   & 0.982  &   0.904 \\[0.1cm]
     \textbf{Weighted Avg.} & 0.789 &   0.157   & 0.818     & 0.789  &  0.755    &  0.661  &  0.952 &    0.929  \\[0.1cm]

	\multicolumn{9}{l}{\cellcolor{gray!10}\textbf{Model 4, RandomCommittee/REPTree, Accuracy = 74.64\%}}\\

     \textbf{LOW} & 0.903  &  0.300   & 0.700  &    0.903  &  0.789  &     0.603  &  0.895 &    0.873 \\[0.1cm]
     \textbf{MEDIUM} & 0.759 &    0.143 &   0.786 &     0.759 &   0.772  &     0.619    & 0.886 &    0.847 \\[0.1cm]
     \textbf{HIGH} & 0.273   & 0.000 &   1.000   &   0.273 &   0.429    &  0.491 &    0.932 &    0.738 \\[0.1cm]
     \textbf{Weighted Avg.} & 0.746 &   0.189  &  0.781  &     0.746 &    0.726     & 0.592  &  0.897  &   0.842  \\[0.1cm]

	\multicolumn{9}{l}{\cellcolor{gray!10}\textbf{Model 5, LWL/LinearNNSearch/DecisionTable, Accuracy = 71.83\%}}\\
	
     \textbf{LOW} & 0.871  &  0.300 &   0.692  &    0.871 &   0.771 &     0.569 &   0.843 &    0.803 \\[0.1cm]
     \textbf{MEDIUM} & 0.759 &    0.190 &   0.733   &   0.759   &  0.746  &    0.565 &   0.800   &  0.729 \\[0.1cm]
     \textbf{HIGH} & 0.182   & 0.000    &  1.000    &  0.182    & 0.308  &    0.398  &  0.823   &  0.541 \\[0.1cm]
     \textbf{Weighted Avg.} & 0.718   & 0.209   &  0.757    &  0.718    & 0.689    &  0.541    & 0.822 &    0.732  \\[0.1cm]

   \noalign{\smallskip}\hline
   \multicolumn{9}{l}{\textbf{TP}-True Positive, \textbf{FP}-False Positive, \textbf{Pre}-Precision, \textbf{Rec}-Recall,}\\
   \multicolumn{9}{l}{  \textbf{F-M}-F-Measure, \textbf{MCC}-Matthews Correlation Coefficient,}\\
   \multicolumn{9}{l}{ \textbf{ROC}-Receiver Operating Characteristic, \textbf{PRC}-Precision-Recall Curve, }\\
   \multicolumn{9}{l}{ \textbf{LOW}-Low level of Cyclomatic Complexity,}\\
   \multicolumn{9}{l}{ \textbf{MEDIUM}-Medium level of Cyclomatic Complexity,}\\
   \multicolumn{9}{l}{ \textbf{HIGH}-High level of Cyclomatic Complexity,}\\
   \multicolumn{9}{l}{ \textbf{W. Avg}-Weighted Average}
\end{tabular}
\end{table}

\textbf{Observation 12: Teams with LOW level of software complexity gains are frequently spotted with higher F-Measure and ROC values.} Our models perform better in detecting subjects achieving low levels of complexity reduction. These are the most critical cases, as such, a software development project manager can quickly detect the teams or individuals responsible for those outcomes and implement actions to bring the project under acceptable quality parameters.

\textbf{Observation 13: Process extended metrics have in general higher importance than process standard metrics.} From Figure \ref{chapter5-b-2.pdf} we can understand that 18 out of 30 metrics are related with the commands issued by the developers. In general, these metrics have also higher importance in the models. It is not surprising to find methods and class extraction commands in the top of the list, with $\approx$86\% and $\approx$56\% importance, respectively. It was however unexpected to find project export actions being so relevant ($\approx$70\%).

\plot{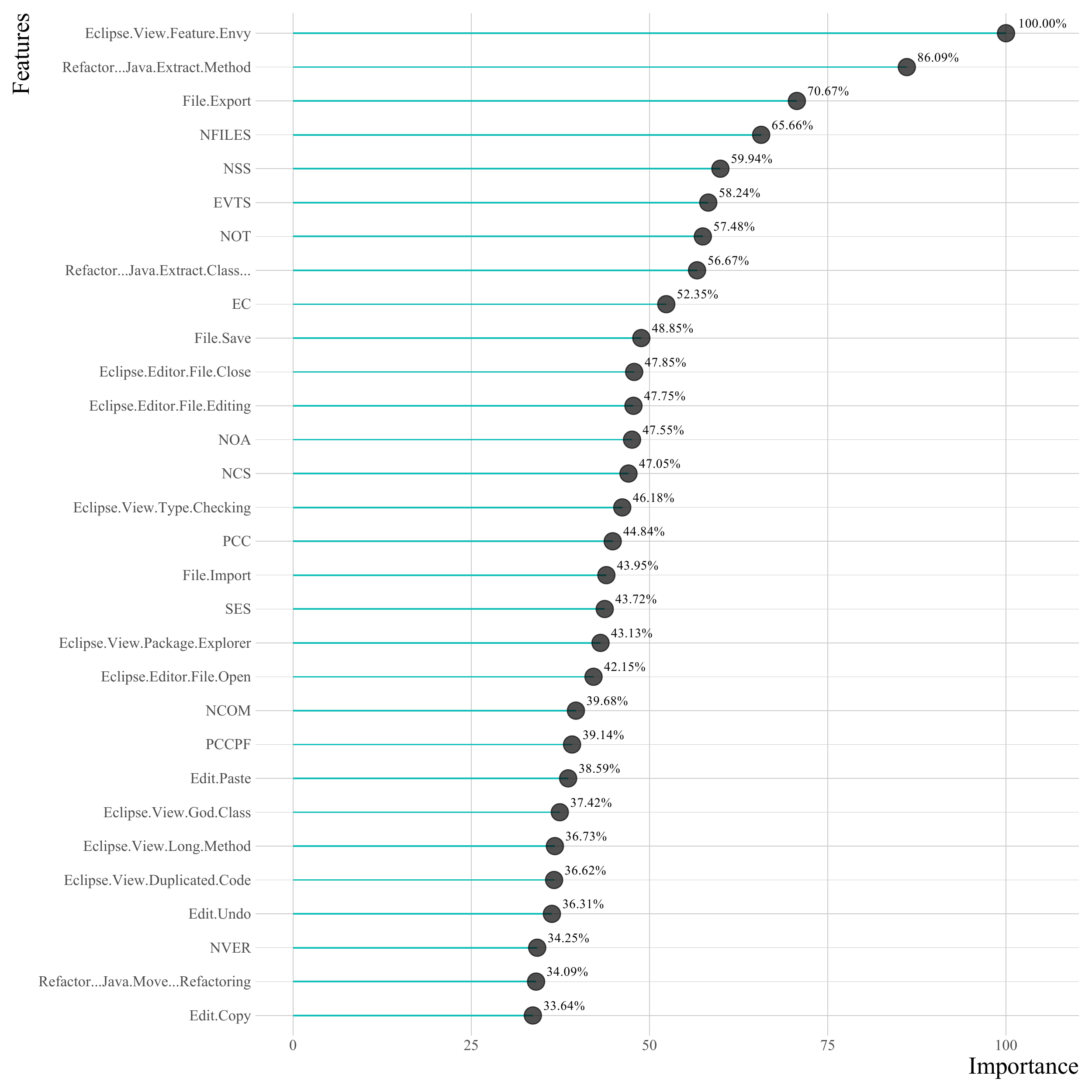}[Feature importance for models on Table \ref{table:model-list2} (Top 30 only)][13.5][H][trim=0cm 0cm 0cm 0cm]

\newpage

\section{Threats to Validity}
\label{sec:Threatsvalidity}

The following types of validity issues were considered when interpreting the results from this article. 

\subsection{Construct Validity}
\label{sec:ConstructValidity}

We acknowledge that this work was supported by an academic environment and using subjects with different maturity and skills which we did not assessed deeply upfront. Additionally, although some work has been done in this domain, we are still just scratching the surface in mining developers' activities using process mining tools. 
Some of these tools are not ready yet to automate the complete flow of: collecting data, discover processes, compute metrics and export results. As a consequence, the referred tasks were mostly done manually, thus, introducing margin for errors in the data metrics used in the experiment. To soften this, and to reduced the risk of having incoherent data, we implemented the validation of metric values from multiple perspectives.
Another possible threat is related with the event data pre-processing tasks before using the Process Mining tools to discover the processes and associated metrics. Events were stored initially in a database, and from there, queries were issued to filter, aggregate and select some process related metrics. We used all the best practices in filtering and querying the data. However, there is always a small chance for the existence of an imprecise query which may have produced incorrect results and therefore, impacted our data analysis.

\subsection{Internal Validity}
\label{sec:InternalValidity}

We used a cluster analysis technique supported by the Elbow and Silhouette methods. This was used to partition the subjects according to different software and process cyclomatic complexity levels. Even if this is a valid approach, other strategies could have been followed, thus, results could vary depending on the alternative methods used, since the models computed to address RQ3 and RQ4 make use of this data partition approach. 
As mentioned earlier, our population was not very large and we had to use it for training and test purposes. As such, our prediction models were all trained using k-fold cross validation and using feature selection methods.

\subsection{External Validity}
\label{sec:ExternalValidity}

We understood from the beginning there was a real possibility that events collected and stored in CSV/JSON files on developers' devices could be manually changed. We tried to mitigate this threat of having data tampering by using a hash function on each event at the moment of their creation. As such, each event contain not only information about the IDE activities, but also a hash code introduced as a new property in the event for later comparison with original event data. For additional precautions regarding data losses, we implemented also real-time event streaming from the IDE to a cloud event hub.

Our initial dataset contains events collected from a group of teams when performing an academic exercise. Each user was provided with a \textit{username} and \textit{password} to enter in the Eclipse Plugin. With this approach, we can easily know which user was working on each part of the software and their role in the whole development process. However, we cannot guarantee that each developer used indeed their own \textit{username}. This does not cause any invalid results in the number of activities for example, but may introduce some bias in the number of developers per team\footnote{A metric used on almost all \textbf{RQs} and identified as having a high importance}.

\subsection{Conclusion Validity}
\label{sec:ConclusionsValidity}

We performed an experiment using data from 71 software teams executing well defined refactoring tasks. This involved 320 sessions of work from 117 developers. Since this is a moderate population size for this type of analysis, we acknowledge this may be a threat to generalize conclusions or make bold assertions.
The Spearmmans' correlation, a nonparametric measure (therefore having less statistical power) of the strength and direction of association that exists between two variables, was done on 32 and 39 teams for automatic and manual refactoring tasks respectively. These figures, although valid, are close to the minimum admissible number of subjects for this type of analysis.
Nevertheless, the insights we unveil in this study should be able to trigger additional research in order to confirm or invalidate our initial findings.

\section{Conclusion}
\label{sec:Conclusions}

\subsection{Main conclusions}

Software maintenance activities, such as refactoring, are said to be very impacted by software complexity. Researchers are measuring software product and processes complexities for a long time and the methods used are frequently debated in the software development realm. However, the comprehension on the links between these two dimensions has a long journey ahead.

In this work, we tried to understand deeper the liaison of process and software complexity. Moreover, we assessed if process driven metrics and IDE issued commands are suitable to build valid models to predict different refactoring methods and/or the expected levels of software cyclomatic complexity variance on development sessions.

We mined the software metrics from a software product after a software quality improvement task was given to a group of developers organized in teams. At the same time we collected events from those developers during the change activities they have performed within the IDE.

To the best of our knowledge, this is the first study where, using proven process mining methods, process metrics were gathered and combined with product metrics in order to understand deeper the liaison of product and process dimensions, particularly the cyclomatic complexities. Furthermore, it brings to the attention of researchers the possibility to adopt process metrics extracted from the IDE usage as a way to complement or even replace product metrics in modeling the development process. 

We can't compare our study to any previous works, however, with a small set of features, we were able to unveil important correlations between product and process dimensions and obtain good models in terms of accuracy and ROC when predicting the type of refactoring done or the expected level of cyclomatic complexity variance after multiple sessions of development. We used a refactoring task as our main use case, however, by taking a snapshot of product and process metrics in different moments in time, one can measure other development practices the same way. 

\subsection{Relevance for practitioners}

This approach can be particularly relevant in cases where product metrics are not available or are difficult to obtain.
It can be also a valid approach to measure and monitor productivity within and between software teams. As we showed by analyzing the sessions complexity and the software cyclomatic complexity variance, non efficient teams can easily be detected.
Our method easily support real-time data collection from individuals located in different geographic zones and with a multitude of development environments. Because the data collection is not dependent on code repositories and is decoupled from check-ins and/or commits, process and code analysis can be performed before repositories are updated. Development organizations can leverage this approach to apply conformance checking methods to verify the adherence of developers' practices with internally prescribed development processes. This facilitates mainly the detection of low performance practices and may trigger quick correction actions from project managers.

\subsection{Limitations}

We are aware that in this work we used only events from the IDE usage. This limits the generalization of the current method. However, our approach, although valid on it's own, may be used to complement project management analysis based on other repositories. Events from tools containing information about the documentation, project management decisions, communication between developers and managers, Q \& A services, test suites and bug tracking systems, together with our method and metrics can build more robust models to comprehend development practices and the relation between software and processes followed to produce it.

\subsection{Future Work}
\label{sec:FutureWork}

In the short term, we plan to apply a similar research approach, but in another context. Instead of refactoring an existing product, we will collect both product and process data for software development from scratch in the context of a programming contest. We intend to assess how the adopted process, both in terms of complexity and efficiency, influences effectiveness, as measured by an automatic judge.

In a previous paper \cite{Caldeira2019AssessingMining} we described how we found that even for a well-defined software development task, there may be a great deal of process variability, due to the human factor. Less focused teams produced more complex process models, due to the spurious/non-essential actions that were carried out and therefore were less efficient. Following this path and using clustering techniques, we expect to derive a catalog of process smells and/or fingerprints to characterize development behaviors. Then, we will use that taxonomy in a personal software process dashboard, that through self-awareness is expected to foster improvement on process efficiency (e.g. less wasted effort) and effectiveness (e.g. yield better deliverables).


We also consider that the following aspects deserve further research efforts:

\begin{itemize}

\item \textbf{Software Repository Diversity.} 
Traditional software repositories have limitations and imprecisions. To expand the analytics coverage on the mining of software development processes, we should explore non trivially used repositories, such as the IDE. This is particularly interesting to drive studies aiming to combine development perspectives: i) product quality and ii) the underlying development process.

\item \textbf{Software Development Process Mining Pipeline.} 
Many process mining tools are not ready for non-human intervention. Due to this reality, many metrics in this article had to be extracted semi-automatically, using a tool but not dispensing user interaction. This is a strong limitation in advancing research based on event data and current process mining methods. A microservices-based architecture seems to be a good alternative for building a coherent pipeline for software development process mining.

\item \textbf{Data Sharing.} 
Research combining software product and process data is scarce and experiments in this area are  difficult to design and execute. To mitigate this problem, we expect an increment in shared datasets containing this hybrid data, providing that privacy and/or anonymity on sensitive information is guaranteed. 

\end{itemize}

\section*{Acknowledgement}
This work was partially funded by the Portuguese Foundation for Science and Technology, under ISTAR-Iscte projects UIDB/04466/2020 and UIDP/04466/ 2020.



\newpage

\bibliography{references,references2}

\newpage


\appendix

\section{Appendix}

\subsection{Product Metrics}
\label{sec:AppendixA1}

\begin{table}[H]
\footnotesize
\caption{Product Metrics Description}
\label{table:product-metrics-description}
\begin{tabular}{p{2cm}lc}
	\hline\noalign{\smallskip}
\textbf{Name} & \textbf{Description} & \textbf{Scale}\\
	\noalign{\smallskip}\hline\noalign{\smallskip}

 \\
\textbf{VG} & McCabe Cyclomatic Complexity (Avg. per Method) & Numeric\\[0.0cm]
\textbf{PAR} & Number of Parameters (Avg. per Method) & Numeric\\[0.0cm]
\textbf{NBD} & Nested Block Depth (Avg. per Method) & Numeric\\[0.0cm]
\textbf{CA} & Afferent Coupling (Avg. per Package Fragment) & Numeric\\[0.0cm]
\textbf{CE} & Efferent Coupling (Avg. per Package Fragment) & Numeric\\[0.0cm]
\textbf{RMI} & Instability (Avg. per Package Fragment) & Numeric\\[0.0cm]
\textbf{RMA} & Abstractness (Avg. per Package Fragment) & Numeric\\[0.0cm]
\textbf{RMD} & Normalized Distance (Avg. per Package Fragment) & Numeric\\[0.0cm]
\textbf{DIT} & Depth of Inheritance Tree (Avg. per Type) & Numeric\\[0.0cm]
\textbf{WMC} & Weighted methods per Class (Avg. per Type) & Numeric\\[0.0cm]
\textbf{NSC} & Number of Children (Avg. per Type) & Numeric\\[0.0cm]
\textbf{NORM} & Number of Overridden Methods (Avg. per Type) & Numeric\\[0.0cm]
\textbf{LCOM} & Lack of Cohesion of Methods (Avg. per Type) & Numeric\\[0.0cm]
\textbf{NOF} & Number of Attributes (Avg. per Type) & Numeric\\[0.0cm]
\textbf{NSF} & Number of Static Attributes (Avg. per Type) & Numeric\\[0.0cm]
\textbf{SIX} & Specialization Index (Avg. per Type) & Numeric\\[0.0cm]
\textbf{NOP} & Number of Packages & Numeric\\[0.0cm]
\textbf{NOC} & Number of Classes (Avg. per Package Fragment) & Numeric\\[0.0cm]
\textbf{NOI} & Number of Interfaces (Avg. per Package Fragment) & Numeric\\[0.0cm]
\textbf{NOM} & Number of Methods (Avg. per Type) & Numeric\\[0.0cm]
\textbf{NSM} & Number of Static Methods (Avg. per Type) & Numeric\\[0.0cm]
\textbf{MLOC} & Method Lines of Code (Avg. per Method) & Numeric\\[0.0cm]
\textbf{TLOC} & Total Lines of Code & Numeric\\
\textbf{TLOC} & Total Lines of Code & Numeric\\
& & \\[0.0cm]
\textbf{VG\_LEVEL} & Different levels of \textbf{\textDelta VG} (\textbf{LOW, MEDIUM, HIGH}) & Categorical\\[0.0cm]

   \noalign{\smallskip}\hline
   
\end{tabular}
\end{table} 

\subsection{Process Metrics}
\label{sec:AppendixA2}

\begin{table}[H]
\footnotesize
\caption{Process Metrics Description}
\label{table:process-metrics-description}
\begin{tabular}{p{2.5cm}lc}
	\hline\noalign{\smallskip}
    \textbf{Name} & \textbf{Description} & \textbf{Scale}\\
	\noalign{\smallskip}\hline\noalign{\smallskip}

\\
\textbf{DEV} & Number of Developers & Numeric\\[0.0cm]
\textbf{SES} & Number of User/Development Sessions & Numeric\\[0.0cm]
\textbf{EVTS} & Number of Events Collected & Numeric\\[0.0cm]
\textbf{NFILES} & Number of Unique Files Touched & Numeric\\[0.0cm]
\textbf{NCOM} & Number of Unique Commands Issued in IDE & Numeric\\[0.0cm]
\textbf{PCCPF} & Process Cyclomatic Complexity per File Touched & Numeric\\[0.0cm]
\textbf{EC} & Number of Event Classes & Numeric\\[0.0cm]
\textbf{NOA} & Number of Activities & Numeric\\[0.0cm]
\textbf{NSS} & Number of Simple States & Numeric\\[0.0cm]
\textbf{NCS} & Number of Composite States & Numeric\\[0.0cm]
\textbf{NOT} & Number of Transitions & Numeric\\[0.0cm]
\textbf{PCC} & Process Cyclomatic Complexity & Numeric\\[0.0cm]
 \textbf{NVER} & Number of Unique IDE Versions & Numeric\\[0.0cm]
 \textbf{NCAT} & Number of Unique Command Categories & Numeric\\[0.0cm]
\textbf{NPLA} & Number of Unique IDE Platforms & Numeric\\[0.0cm]
\textbf{NISP} & Number of Unique Geographic Locations & Numeric\\[0.0cm]
\textbf{NOS} & Number of Unique Operating Systems & Numeric\\[0.0cm]
 \textbf{NPER} & Number of Unique Perspectives used in the IDE & Numeric\\[0.0cm]
& & \\
\textbf{PCC\_LEVEL} & Different levels of \textbf{PCC} (\textbf{LOW, HIGH}) & Categorical\\[0.2cm]

   \noalign{\smallskip}\hline
\end{tabular}
\end{table}

\begin{table}[H]
\footnotesize
\caption{Process-Extended Metrics Description}
\label{table:process-extended-metrics-description}
\begin{tabular}{p{3cm}p{6.4cm}c}
	\hline\noalign{\smallskip}
	\textbf{Category} & \textbf{Name} & \textbf{Scale}\\
	\noalign{\smallskip}\hline\noalign{\smallskip}

\multirow{8}{*}{\textbf{Refactor}} 
& Java-Extract Method & Numeric\\[0.0cm]
& Java-Move - Refactoring & Numeric\\[0.0cm]
& Java-Extract Class... & Numeric\\[0.0cm]
& Java-Rename - Refactoring & Numeric\\[0.0cm]
& Delete Resources & Numeric\\[0.0cm]
& Java-Encapsulate Field & Numeric\\[0.0cm]
& Java-Change Method Signature & Numeric\\[0.0cm]
& Java-Move Type to New File & Numeric\\[0.1cm]
\hline
\multirow{3}{*}{\textbf{Eclipse Editor}}
& File Open & Numeric\\[0.0cm]
& File Editing & Numeric\\[0.0cm]
& File Close & Numeric\\[0.1cm]
\hline
\multirow{12}{*}{\textbf{Eclipse View}}
& Project Explorer & Numeric\\[0.0cm]
& Package Explorer & Numeric\\[0.0cm]
& Long Method & Numeric\\[0.0cm]
& God Class  & Numeric\\[0.0cm]
& Code Smell Visualization  & Numeric\\[0.0cm]
& Type Checking  & Numeric\\[0.0cm]
& Feature Envy  & Numeric\\[0.0cm]
& Duplicated Code  & Numeric\\[0.0cm]
%
\hline
\multirow{7}{*}{\textbf{Edit}}
& Find and Replace  & Numeric\\[0.0cm]
& Copy  & Numeric\\[0.0cm]
& Paste  & Numeric\\[0.0cm]
& Cut  & Numeric\\[0.0cm]
& Delete  & Numeric\\[0.0cm]
& Undo  & Numeric\\[0.0cm]
& Redo  & Numeric\\[0.1cm]
\hline
\multirow{4}{*}{\textbf{File}}
& Import  & Numeric\\[0.0cm]
& Refresh  & Numeric\\[0.0cm]
& Save  & Numeric\\[0.0cm]
& Save All  & Numeric\\[0.1cm]
\hline
\textbf{Source} & Generate Getters and Setters  & Numeric\\[0.0cm]
\textbf{Compare} & Select Next Change  & Numeric\\[0.2cm]
....... & //List is truncated on purpose  & \\[0.0cm]
....... & //List size is $\approx$250  &\\[0.2cm]
\textbf{Text Editing} & Delete Previous Word   & Numeric\\[0.1cm]
%

   \noalign{\smallskip}\hline
\end{tabular}
\end{table}

\newpage

\subsection{Algorithms shown in Model Evaluations}
\label{sec:AppendixA3}

\footnotesize
\textbf{RandomCommittee.}
Method for building an ensemble of randomizable base classifiers. Each base classifier is built using a different random seed number (but based one the same data). The final prediction is a straight average of the predictions generated by the individual base classifiers.\\

\noindent\textbf{RandomSubSpace.}
This method constructs a decision tree based classifier that maintains highest accuracy on training data and improves on generalization accuracy as it grows in complexity. The classifier consists of multiple trees constructed systematically by pseudo-randomly selecting subsets of components of the feature vector, that is, trees constructed in randomly chosen sub-spaces.\\

\noindent\textbf{RandomForest.}
Method for constructing a forest of random trees. It consists of a learning method for classification, regression and other tasks that operates by constructing a multitude of decision trees at training time and outputting the class that is the mode of the classes (classification) or mean prediction (regression) of the individual trees.\\

\noindent\textbf{RepTree.}
Fast decision tree learner. Builds a decision/regression tree using information gain/variance and prunes it using reduced-error pruning (with back-fitting). Only sorts values for numeric attributes once. Missing values are dealt with by splitting the corresponding instances into pieces.\\

\noindent\textbf{LMT.}
'Logistic Model Trees' are classification trees with logistic regression functions at the leaves. The algorithm can deal with binary and multi-class target variables, numeric and nominal attributes and missing values.\\

\noindent\textbf{Logistic Regression.}
Method for building and using a multinomial logistic regression model with a ridge estimator.
Logistic regression is a statistical model that in its basic form uses a logistic function to model a binary dependent variable, although more complex extensions exist.\\

\noindent\textbf{LWL.}
The Locally Weighted Learning method uses an instance-based algorithm to assign instance weights which are then used by a specified WeightedInstancesHandler. Can do classification (e.g. using naive Bayes) or regression (e.g. using linear regression).\\

\noindent\textbf{LinearNNSearch.}
This method implements the brute force search algorithm for nearest neighbour search.\\

\noindent\textbf{DecisionTable.}
Builds and uses a simple decision table majority classifier.\\

\noindent\textbf{Bagging.}
Method for bagging a classifier to reduce variance. Can do classification and regression depending on the base learner.\\

\noindent\textbf{KStar.}
Is an instance-based classifier, that is, the class of a test instance is based upon the class of those training instances similar to it, as determined by some similarity function.  It differs from other instance-based learners in that it uses an entropy-based distance function.\\

\newpage

\subsection{Best-Fit Models - Source Code}
\label{sec:AppendixA4}

\scriptsize
\lstset{
    tabsize = 4, 
    showstringspaces = false, 
    numbers = left, 
    keywordstyle = \color{iscte-iul-palette}, 
    stringstyle = \color{black}, 
    rulecolor = \color{black}, 
    breaklines = true, 
    numberstyle = \small, 
    postbreak=\mbox{\textcolor{iscte-iul-palette}{$\hookrightarrow$}\space}
}
\begin{lstlisting}[language=java, caption=Best-Fit Model Code for Refactoring Practice Detection, label=lst:weka-best-model]
/** Java code to implement the best model found. */
 
/** Attribute Search **/
AttributeSelection as = new AttributeSelection();
ASSearch asSearch = ASSearch.forName("weka.attributeSelection.GreedyStepwise", new String[]{"-C", "-R"});
as.setSearch(asSearch);

/** Attribute Evaluation and Selection **/
ASEvaluation asEval = ASEvaluation.forName("weka.attributeSelection.CfsSubsetEval", new String[]{"-M", "-L"});
as.setEvaluator(asEval);
as.SelectAttributes(instances);

/** Reduce Dimensions **/
instances = as.reduceDimensionality(instances); 

/** Build Classifier **/
Classifier classifier = AbstractClassifier.forName("weka.classifiers.meta.RandomCommittee", new String[]{"-I", "64", "-S", "1", "-W", "weka.classifiers.trees.RandomForest", "--", "-I", "29", "-K", "13", "-depth", "3"});
classifier.buildClassifier(instances);
\end{lstlisting}

\lstset{
    tabsize = 4, 
    showstringspaces = false, 
    numbers = left, 
    keywordstyle = \color{iscte-iul-palette}, 
    stringstyle = \color{black}, 
    rulecolor = \color{black}, 
    breaklines = true, 
    numberstyle = \small, 
    postbreak=\mbox{\textcolor{iscte-iul-palette}{$\hookrightarrow$}\space}
}
\begin{lstlisting}[language=java, caption=Best-Fit Model Code for expected Cyclomatic Complexity level detection, label=lst:weka-best-model2]
/** Java code to implement the best model found. */

/** Attribute Search **/
AttributeSelection as = new AttributeSelection();
ASSearch asSearch = ASSearch.forName("weka.attributeSelection.GreedyStepwise", new String[]{"-C", "-R"});
as.setSearch(asSearch);

/** Attribute Evaluation and Selection **/
ASEvaluation asEval = ASEvaluation.forName("weka.attributeSelection.CfsSubsetEval", new String[]{"-L"});
as.setEvaluator(asEval);
as.SelectAttributes(instances);

/** Reduce Dimensions **/
instances = as.reduceDimensionality(instances);

/** Build Classifier **/
Classifier classifier = AbstractClassifier.forName("weka.classifiers.lazy.LWL", new String[]{"-K", "60", "-A", "weka.core.neighboursearch.LinearNNSearch", "-W", "weka.classifiers.rules.DecisionTable", "--", "-E", "auc", "-S", "weka.attributeSelection.GreedyStepwise", "-X", "2"});
classifier.buildClassifier(instances);



\end{lstlisting}

\newpage




\end{document}